\documentclass[floats,floatfix,aps,amsfonts,notitlepage,superscriptaddress,twocolumn,eqsecnum,nofootinbib]{revtex4-1}
\usepackage{ifpdf}
\usepackage[utf8]{inputenc}
\usepackage[UKenglish]{babel}
\usepackage{graphicx}
\usepackage[colorlinks]{hyperref}
\usepackage{amsmath, amsthm, amssymb}
\usepackage{multirow}
\usepackage{url}
\usepackage{subfigure}
\usepackage{color}
\usepackage[usenames,dvipsnames,svgnames,table]{xcolor}
\usepackage{tabularx}

\allowdisplaybreaks[1]

\definecolor{CiteColor}{rgb}{0,0.5,0}
\hypersetup{citecolor=CiteColor}
\definecolor{RefColor}{rgb}{0.55,0,0}
\hypersetup{linkcolor=RefColor}
\definecolor{darkgreen}{rgb}{0.2,0.7,0.2}

\newcommand{\diff}[2]  {\frac{d #1}{d #2}}

\newcommand{\en}{E}
\newcommand{\ang}{L}

\newcommand{\self}{\text{self}}

\newcommand{\eff}{\text{eff}}
\newcommand{\ret}{\text{ret}}
\newcommand{\res}{\text{res}}
\newcommand{\win}{\mathcal{W}}
\newcommand{\bh}[1]{\bar{h}^{(#1)}}

\newcommand{\hb}[1]{\bar{h}^{(#1)}}
\newcommand{\hbP}[1]{\bar{h}^{(#1)P}}
\newcommand{\hbm}[1]{\tilde{h}^{(#1)-}}
\newcommand{\hbp}[1]{\tilde{h}^{(#1)+}}
\newcommand{\hbpm}[1]{\tilde{h}^{(#1)\pm}}
\newcommand{\hbres}[1]{\bar{h}^{(#1)\text{res}}}

\newcommand{\ab}{{\bar{a}}}
\newcommand{\bb}{{\bar{b}}}

\newcommand{\sbar}{{\bar{s}}}

\newcommand{\ellh}{\hat{\ell}}

\renewcommand{\c}{\cos \omega_r \tau}

\newcommand{\f}{\frac}

\newcommand{\nn}{\nonumber}

\newcommand{\be}{\begin{equation}}
\newcommand{\ee}{\end{equation}}
\newcommand{\ba}{\begin{eqnarray}}
\newcommand{\ea}{\end{eqnarray}}

\renewcommand{\c}{\hskip0.05cm,}
\newcommand{\p}{\hskip0.05cm.}

\mathchardef\mhyphen="2D

\def\prd{Phys. Rev. D}
\def\apj{Ap. J.}
\def\etal{\textit{et al.}}

\def\half{\frac{1}{2}}

\begin{document}

\title{Applying the effective-source approach to frequency-domain self-force calculations: Lorenz-gauge gravitational perturbations}
\author{Barry Wardell}
\affiliation{Department of Astronomy, Cornell University, Ithaca, NY 14853, USA}
\affiliation{School of Mathematical Sciences and Complex \& Adaptive Systems Laboratory, University College Dublin, Belfield, Dublin 4, Ireland}
\author{Niels Warburton}
\affiliation{MIT Kavli Institute for Astrophysics and Space Research, Massachusetts Institute of Technology, Cambridge, MA 02139, USA}
\affiliation{School of Mathematical Sciences and Complex \& Adaptive Systems Laboratory, University College Dublin, Belfield, Dublin 4, Ireland}

\begin{abstract}
	With a view to developing a formalism that will be applicable at second
	perturbative order, we devise a new practical scheme for computing the
	gravitational self-force experienced by a point
	mass moving in a curved background spacetime. Our method works in the
	frequency domain and employs the effective-source approach, in which a
	distributional source for the retarded metric perturbation is replaced with an
	effective source for a certain regularized self-field. A key ingredient of the calculation is the analytic determination of an
	appropriate puncture field from which the effective source and regularized
	residual field can be calculated. In addition to its application in our
	effective-source method, we also show how this puncture field can be used to derive
	tensor-harmonic mode-sum regularization parameters that improve the
	efficiency of the traditional mode-sum procedure. To demonstrate the method, we calculate the first-order-in-the-mass-ratio self-force and redshift
	invariant for a point mass on a circular orbit in Schwarzschild spacetime.
\end{abstract}

\maketitle

\section{Introduction} 

The detection of gravitational waves from inspiraling compact binary systems
will be greatly assisted by theoretical waveform templates. Using matched
filtering techniques, these templates will help to extract the incoming signal
from the detector noise and, once detection becomes routine, the same templates
will allow the parameters of the source binary systems to be established.
Constructing an appropriate bank of waveform templates necessitates
understanding the two-body problem in the general-relativistic context. Unlike
its Newtonian counterpart, this problem cannot be solved analytically with
closed-form solutions. Instead, a variety of methods exist to approximate the
solution, each best suited to a particular set of system parameters.

To model binary systems of comparable mass and small orbital separation it is
necessary to turn to numerical simulations to solve the full non-linear
Einstein field equations. Numerical Relativity has made great progress in
recent years and it is now routine to numerically evolve binary systems of
comparable-mass black holes or neutron stars through tens of orbits up to, and
through, merger \cite{Mroue_etal:2013,Bernuzzi_etal:2015}. Aside from numerical truncation error, these results are
exact. However, the computational cost of running simulations to merger
increases rapidly as the mass ratio of the smaller to the more massive body is
decreased. The computational cost similarly increases as the initial orbital
separation increases. Consequently, in these domains other approaches are
required.

For systems with large orbital separation the post-Newtonian (PN) expansion can
be employed. This is a perturbative approach to the problem that involves
expanding the field equations in powers of the orbital velocity as a fraction
of the speed of light. This approach has a long and rich history and today the
post-Newtonian expansion of the dynamics of a binary is now known up to 3PN order, with many 4PN terms now known --- see
Ref.~\cite{Blanchet:living_review} for a recent review.

When one of the bodies is substantially more massive than the other the small
mass ratio of the system can be used as a perturbative parameter. In this
approach the less massive of the two bodies is usually modeled as a point
particle. Flux balance arguments allow the dissipative dynamics to be modeled \cite{Drasco-Hughes,Fujita-Hikida-Tagoshi},
but in order to include conservative corrections it is necessary to evaluate
the local `self-force' acting on the particle. With a point particle source
this then necessitates a regularization procedure to remove the Coulomb-like
divergence of the metric perturbation that does not contribute to the orbital
dynamics. Over the years this regularization procedure has been placed on very
firm theoretical footing at first order in the mass ratio
\cite{Mino-Sasaki-Tanaka,Quinn-Wald,Poisson-review} and recently has been
understood at second perturbative order
\cite{Pound:2nd_order,Gralla:2nd_order,Detweiler:2nd_order}.

At first perturbative order a large number of practical self-force calculations
have been made -- see Ref.~\cite{Wardell:review} for a recent review. Practical
calculation techniques are often prototyped with scalar-field toy models before
the gravitational case is considered. In both cases motion in the spherically
symmetric Schwarzschild spacetime is usually tackled first, before turning
attention to motion in the more astrophysically relevant Kerr spacetime of a
rotating black hole.

In recent years, the calculation of gauge-invariant quantities has proven to be
particularly fruitful as it allows for comparison of self-force results with
those of post-Newtonian theory and Numerical Relativity. A number of these
gauge-invariant quantities are now known
\cite{Detweiler-circular,Dolan_etal:spin_invariant,Dolan_etal:tidal_invariants,Nolan_etal:2015}
and using them to make cross-cultural comparisons has been illuminating,
helping to delineate the region in which perturbative approaches are valid, as
well as acting as a cross-check on the widely different computational approaches
taken by each scheme. One particularly interesting result is that of Le Tiec
\etal~\cite{LeTiec_etal-periastron_advance} showing that even for comparable
mass systems perturbative methods can make meaningful statements about the
orbital dynamics. Calibrating Effective-One-Body theory is another important use
for gauge-invariant results
\cite{Akcay_etal:2012,Bini-Damour:spin,Bini-Damour:tidal}.

When making self-force calculations, an important consideration is the practical
regularization technique to be employed. All the techniques derive from the
same fundamental regularization procedure, but different techniques suit
different calculational approaches. One of the most commonly employed techniques is the mode-sum prescription \cite{mode-sum-orig}. This approach
relies on the mode-decomposition rendering the individual modes of the retarded
field solution finite at the particle's location and has had much success with
1+1 time-domain and frequency-domain calculations. For 2+1 or 3+1 time-domain
decompositions, where the retarded field remains divergent at the particle, so
called `effective source' techniques were developed \cite{Barack-Golbourn,
Vega-Detweiler}. This approach involves moving the contribution from the
singular metric perturbation near the particle into the source. This procedure
renders the otherwise divergent source finite and amenable to numerical
treatment.

To date all self-force calculations have been at first order in the mass ratio.
The key motivation for the present work is to develop a set of techniques that
can extend the work mentioned above to encompass second-order-in-the-mass-ratio
calculations, with the aim of computing conservative gauge-invariant quantities
\cite{Pound:2014}. This immediately suggests the basic form our approach should
take:
\begin{enumerate}
  \item Work in the Lorenz gauge. Currently the regularization procedure at
        second order in the mass ratio is best understood in the Lorenz-gauge
        \cite{Pound:2nd_order,Pound-Miller:2014}.
  \item Work in the frequency domain. At present it is not known how to stably
        evolve the monopole and dipole contributions to the Lorenz gauge
        linearized Einstein equation \cite{Dolan-Barack:GSF_m_mode}. These instabilities are observed in
        $1+1$D, $2+1$D and $3+1$D time-domain decompositions on a Schwarzschild background,
        and similar instabilities have been seen in Kerr spacetime \cite{Dolan:Capra16}. Even if the
        multipole $l\ge2$ modes are evolved separately in the time domain, the
        $\ell=0,1$ modes must be solved in the frequency domain.
  \item Regularize with an effective source. Unlike at first order in the mass
        ratio, the individual multipole modes of the second-order retarded field
        diverge at the particle's location. This precludes the use of the
        mode-sum method for regularization as it requires the multipole modes of
        the retarded metric perturbation to be finite at the particle's location.
\end{enumerate}
The details of points 2 and 3 were fleshed out by the authors in
Ref.~\cite{Warburton-Wardell} using a toy scalar-field example. In this work we
address point 1 and extend our previous scalar-field results to cover
Lorenz-gauge gravitational perturbations generated by a point mass on a circular
orbit in a Schwarzschild background spacetime. We leave for future work the
issues of extending our approach to eccentric orbits (possibly making use of
methods developed for the conventional mode-sum scheme \cite{Hopper:2015jxa})
and to a Kerr background spacetime (which would likely require more extensive
modification due to the common practice of using radiation gauge and spin-weighted
\emph{spheroidal} harmonics in order to achieve separability of the field equations
\cite{Teukolsky,Teukolsky2}).

In addition to laying out a formalism that will be applicable at
second order in the mass ratio, a natural by-product of this work is the
extension of the standard mode-sum scheme to allow for the direct regularization of
the retarded tensor modes of the metric perturbation. In the standard mode-sum
procedure the retarded tensor modes of the metric perturbation must be
projected onto a basis of scalar spherical-harmonics before regularization can
be performed. This step is cumbersome and, due to the coupling between scalar
and tensor modes, requires the computation of additional tensor modes. Our
new prescription neatly avoids these issues altogether.

The layout of this article is as follows. In
Sec.~\ref{sec:field_eqs_and_ret_sol} we overview the Lorenz-gauge field
equations and their retarded solution for circular orbits in a background Schwarzschild
geometry. In Sec.~\ref{sec:regularization} we discuss the standard mode-sum
and effective-source regularization procedures. In Sec.~\ref{sec:Seff_in_FD} we
detail the construction of an effective source for a Lorenz-gauge metric
perturbation sourced by a particle on a circular orbit. In
Sec.~\ref{sec:num_implementation_and_results} we outline our numerical
procedure and give our results. Finally, in Sec.~\ref{sec:mode_sum_tensor} we
extend the mode-sum prescription to work directly with tensor-harmonic modes. There is
additional supporting material in Appendices
\ref{sec:harmonics}-\ref{apdx:monopole}.

This paper follows the conventions of Misner, Thorne and Wheeler
\cite{Misner:1974qy}; a ``mostly positive'' metric signature, $(-,+,+,+)$, is
used for the spacetime metric, the connection coefficients are defined by
$\Gamma^{\lambda}_{\mu\nu}=\frac{1}{2}g^{\lambda\sigma}(g_{\sigma\mu,\nu}
+g_{\sigma\nu,\mu}-g_{\mu\nu,\sigma}$), the Riemann tensor is
$R^{\tau}{}_{\!\lambda\mu\nu}=\Gamma^{\tau}_{\lambda\nu,\mu}
-\Gamma^{\tau}_{\lambda\mu,\nu}+\Gamma^{\tau}_{\sigma\mu}\Gamma^{\sigma}_{\lambda\nu} -\Gamma^{\tau}_{\sigma\nu}\Gamma^{\sigma}_{\lambda\mu}$, the Ricci
tensor and scalar are $R_{\mu\nu}=R^{\tau}{}_{\!\mu\tau\nu}$ and
$R=R_{\mu}{}^{\!\mu}$, and the Einstein equations are
$G_{\mu\nu}=R_{\mu\nu}-\frac{1}{2}g_{\mu\nu}R=8\pi
T_{\mu\nu}$. Standard geometrized units are used, with $c=G=1$. Greek
indices are used for four-dimensional spacetime components and capital Latin
letters are used for indices on the two-sphere. We work with standard
Schwarzschild coordinates $(t,r,\theta,\varphi)$ and also work with a second
coordinate system, $(t,r,\alpha,\beta)$, which is related by a rotation. We
denote a point on the worldline of the point mass by a `0' subscript
(e.g., $x_0$) and indices on tensors evaluated on the worldline are indicated
by an overbar (e.g. $g_{\bar{\mu}\bar{\nu}}$). We also find it
useful to define $f\equiv 1-2M/r$ where $M$ is the mass of the background
Schwarzschild black hole.

\section{Lorenz-gauge field equations and their retarded solution for a circular orbit}\label{sec:field_eqs_and_ret_sol}

In this section we overview the Lorenz-gauge field equations for
linear-in-the-mass-ratio perturbations of Schwarzschild spacetime, along with
their decomposition into tensor-harmonic and frequency modes. The basis of
tensor-harmonic modes we use is that of Barack and Sago \cite{Barack-Sago-circular,Barack-Sago-eccentric}, themselves a modification of the basis given by Barack and Lousto \cite{Barack-Lousto-2005}. Barack and Sago also further
decomposed the monopole and dipole field equations into the frequency-domain to
side-step instabilities that occur when evolving those modes in the time-domain
(see Ref.~\cite{Dolan-Barack:GSF_m_mode} for a discussion of this issue).
Later, the frequency decomposition was given for all modes with calculations
for generic bound orbits by Akcay
\etal~\cite{Akcay-GSF-circular,Akcay-Warburton-Barack} and Osburn
\etal~\cite{Osburn_etal:2014}.

\subsection{Field Equations}

Let us denote by $g$ the full spacetime metric, which we shall consider to be
the sum of the metric perturbation, $h$, and the background Schwarzschild
metric, $\mathring{g}$, such that $g = \mathring{g} + h$. Hereafter an over-ring
will denote a quantity defined with respect to the background (vacuum)
spacetime. In a given coordinate system, the Einstein field equations will then
take the form
\begin{align}
	G_{\mu\nu}[\mathring{g}_{\mu\nu}+h_{\mu\nu}] = 8\pi T_{\mu\nu}
\end{align}
where $G$ is the Einstein tensor, a functional of the full spacetime metric
$g$, and $T$ is the stress-energy tensor. Let us define the trace of the metric
perturbation by $\text{Tr}(h)=\mathring{g}^{\mu\nu}h_{\mu\nu}$. We shall find
that the field equations for the metric perturbation take a simpler form
when expressed in terms of the trace-reversed metric perturbation,
$\bar{h}_{\mu\nu}$, defined by
\begin{equation}
	\bar{h}_{\mu\nu} \equiv h_{\mu\nu} - \frac{1}{2}\mathring{g}_{\mu\nu}\text{Tr}(h)
\end{equation}
so named because $\text{Tr}(\bar{h})=-\text{Tr}(h)$. 

As discussed in the Introduction, when we approach a
second-order-in-the-mass-ratio calculation we will want to work in the
Lorenz gauge. Consequently, to develop the necessary techniques we will work in
the Lorenz-gauge with the first-order-in-the-mass-ratio calculation we present
in this work. The Lorenz gauge condition is defined by
\begin{equation}\label{eq:lorenz_gauge}
	\mathring{\nabla}_\mu \bar{h}^{\mu\nu}=0 \c
\end{equation}
where the covariant derivative is taken with respect to the background metric.
By expanding the Einstein tensor in powers of the mass ratio and only retaining
terms linear in $\mu$ we arrive at the (Lorenz-gauge) linearized Einstein
equation given by
\begin{eqnarray}\label{eq:linearized_einstein}
	\mathring{\square}\bar{h}_{\mu\nu} + 2 \mathring{R}^{\rho\;\;\sigma}_{\;\;\mu\;\;\nu}\bar{h}_{\rho\sigma} = -16\pi T_{\mu\nu}
\end{eqnarray}
where $\mathring{\square} = \mathring{\nabla}_\mu\mathring{\nabla}^\mu$ and $\mathring{R}$
is the Riemann tensor of the background spacetime. In this work we shall take
the metric perturbation to be sourced by a point particle of mass $\mu$. The
corresponding energy-momentum tensor is given by
\begin{equation}\label{eq:GSF_source}
	T_{\mu\nu} = \mu\int^\infty_{-\infty} [-\det(\mathring{g})]^{-1/2} \delta^4(x^\mu - x_0^\mu)u_\mu u_\nu\, d\tau		\c
\end{equation}
where $\det(\mathring{g})=r^4\sin^2\theta$ is the determinant of the background
metric tensor, $u^\mu$ is the particle's four-velocity and $\tau$ is the proper
time measured along the particle's worldline. We also use $x^\mu$ to denote a
general spacetime coordinate and hereafter adopt the notation that a subscript
`$0$' denotes a quantity's value evaluated at the particle. Note that for a circular orbit in the equatorial plane of a Schwarzschild black hole we have $u_r=u_\theta = 0$ and $u_t=-\en_0$, $u_\varphi = \ang_0$ where
\begin{align}
	\en_0 = f_0\sqrt{\frac{r_0}{r_0-3M}},\qquad \ang_0 = r_0\sqrt{\frac{M}{r_0-3M}},
\end{align}
are the (specific) orbital energy and angular momentum, respectively. Finally, we 
mention that the gauge equation \eqref{eq:lorenz_gauge} and field equation \eqref{eq:linearized_einstein} are
consistent so long as the particle is moving along a geodesic of the background
spacetime (as then $\nabla_\mu T^{\mu\nu} = 0$).

The field equation in the form of Eq.~\eqref{eq:linearized_einstein} is not
well suited to a numerical treatment as the metric perturbation diverges in the
vicinity of the worldline. Instead, an effective-source approach can be
employed to regularize the field equation and allow for a certain regular field
to be solved for directly. Alternatively, with a $1+1$D or frequency-domain
decomposition the individual multipole modes of the metric perturbation become
finite at the particle's location and the mode-sum scheme can be employed to
regularize on a mode-by-mode basis --- see Sec.~\ref{sec:regularization} below
for an overview of these two regularization procedures. As discussed in the
Introduction, at second order it will become necessary to employ an
effective-source scheme even within a multipole decomposition. For that reason,
despite not being required for a first-order-in-the-mass-ratio calculation, we will pursue an
effective-source approach within a multipole and Fourier decomposition of the
field equations. We give the details of this decomposition now.

\subsection{Decomposition into tensor-harmonic and frequency modes}

In this section we overview the multipole and Fourier decomposition of the
metric perturbation and source. The explicit details of this decomposition have
been laid out elsewhere \cite{Barack-Lousto,Akcay-GSF-circular} and are
summarized in Appendix~\ref{sec:harmonics}; here we shall just present the
key results required for this work.

There are many different conventions and notations used to define a
tensor-harmonic basis. In this work we use the definition chosen by Barack and
Lousto \cite{Barack-Lousto} with the slight modification introduced by Barack
and Sago \cite{Barack-Sago-circular}. The key property of the
Barack-Lousto-Sago tensor-harmonics is that they form a ten-dimensional basis
for any second rank, symmetric 4-dimensional tensor field in Schwarzschild
spacetime. This allows us to write the ten independent components of the
(trace-reversed) metric perturbation in terms of the spherical-harmonic modes
of ten fields, $\bar{h}^{(i)}_{\ell m}(t,r)$ for $i=1, \ldots, 10$ via
\begin{equation} \label{eq:hilm-tr}
  \bar{h}^{(i)}_{\ell m}(t,r) = \frac{r}{\mu\, a^{(i)}_\ell} \int_0^{2\pi} \int_0^\pi \bar{h}_{\tau\kappa} \eta^{\tau\mu}\eta^{\kappa\nu} Y_{\mu\nu}^{(i)\ell m}{}^\ast\, d\Omega
\end{equation} 
where $d\Omega=\sin\theta\,d\theta\,d\varphi$ and the details of the tensor
basis (including definitions for $Y_{\mu\nu}^{(i)\ell m}$, $\eta^{\mu\nu}$ and $a^{(i)}_\ell$)
are given in Appendix~\ref{sec:harmonics}. We further decompose into
Fourier-frequency modes,
\begin{equation}
 \bar{h}^{(i)}_{\ell m} (t, r) \equiv \frac{1}{2\pi} \int_{-\infty}^{\infty} \bar{h}^{(i)}_{\ell m} (\omega, r) e^{-i \omega t} dt.
\end{equation}
For periodic motion the integral over frequencies reduces to a sum over
discrete harmonics (hereafter modes). In particular, for a circular geodesic
orbit the mode frequency is a simple overtone of the fundamental azimuthal
frequency
\begin{align}
	\omega_m = m \Omega_\varphi,
\end{align}
where $\Omega_\varphi \equiv d\varphi_0/dt = \sqrt{M/r_0^3}$ and $m$ is the
azimuthal mode index. In the remainder of this article, we can therefore denote
the modes by $\bar{h}^{(i)}_{\ell m}(r)$ without any ambiguity. The
stress-energy tensor can be similarly decomposed into modes $T_{\ell m}^{(i)}(r)$ \cite{Barack-Lousto-2005}.

Substituting the mode expansions of $\bar{h}_{\mu\nu}$ and $T_{\mu\nu}$ into
the linearized Einstein equation, \eqref{eq:linearized_einstein}, the
angular and time dependence decouples. The spherical symmetry of the background
geometry ensures that the individual multipole modes are eigenfunctions of the
wave operator \eqref{eq:linearized_einstein} and consequently each multipole
mode can be solved for independently from the others, though in general the ten
tensorial components of each mode remain coupled. The resulting set of ordinary
differential equations for each multipole mode are given by
\begin{eqnarray}\label{eq:metric_pert_FD}
  \square^{sc}_{\ell m} \bar{h}_{\ell m}^{(i)} - 4f^{-2}\mathcal{M}^{(i)}{}_{(j)} \bar{h}_{\ell m}^{(j)} = \mathcal{J}_{\ell m}^{(i)}\delta(r-r_0),
\end{eqnarray}
where $\mathcal{J}^{(i)}_{\ell m}$ comes from the decomposition of the source and
$\square^{sc}_{\ell m}$ is the scalar wave operator,
\begin{eqnarray}
  \square^{sc}_{\ell m} =  \diff{}{r^2} + \frac{f'}{f}\diff{}{r} -  f^{-2} \left[V_{\ell}(r) - \omega_m^2\right].
\end{eqnarray}
Here, a prime denotes differentiation with respect to $r$ and the potential term
is given by
\begin{equation}
  V_{\ell}(r) = f(r)\left[\frac{2M}{r^3} + \frac{\ell(\ell+1)}{r^2}\right].
\end{equation}
The $\mathcal{M}^{(i)}{}_{(j)}$ that appear in Eq.~\eqref{eq:metric_pert_FD} are
first-order differential operators that couple the ten components of the metric
perturbation; we give their explicit form in
Appendix~\ref{apdx:field_eqs_coupling}. In deriving the
$\mathcal{M}^{(i)}{}_{(j)}$ we use in this work, we have used the frequency-domain
decomposition of the Lorenz-gauge condition \eqref{eq:lorenz_gauge} to simplify
the resulting equations. This decomposition of the Lorenz-gauge condition is
given by
\begin{align}
i\omega_m\hb{1} =& -f\left(i\omega_m\hb{3} + \hb{2}_{,r} + \frac{\hb{2} - \hb{4}}{r}\right), \label{eq:gauge_even1}\\
i\omega_m\hb{2} =& -f\hb{1}_{,r} + f^2\hb{3}_{,r}\nonumber \\
&-\frac{f}{r}\left(\hb{1}-\hb{5}-f\hb{3}-2f\hb{6}\right), \label{eq:gauge_even2}\\
i\omega_m\hb{4} =& - \frac{f}{r}\left(r\hb{5}_{,r} + 2\hb{5} + L\hb{6} - \hb{7}\right), \label{eq:gauge_even3}\\
i\omega_m\hb{8} =&- \frac{f}{r}\left(r \hb{9}_{,r} + 2\hb{9} - \hb{10}\right), \label{eq:gauge_odd}
\end{align}
where $L=\ell(\ell+1)$.

The ten field equations \eqref{eq:metric_pert_FD} are not all coupled together;
instead they separate out into independent even- ($i=1,\ldots,7$) and odd-parity
($i=8,9,10$) sectors. Examining the sources (given in
Appendix~\ref{apdx:field_eqs_coupling}) we see that
\begin{align}
  \mathcal{J}^{(i=1,\dots,7)} &\propto [Y^{\ell m}(\pi/2, \Omega_\varphi t)]^* = 0\quad \text{for}\;\; \ell+m = \text{odd}  \nonumber  \\
  \mathcal{J}^{(i=8,9,10)} &\propto [Y^{\ell m}_{,\theta}(\pi/2, \Omega_\varphi t)]^* = 0\quad \text{for}\;\; \ell+m = \text{even}  \nonumber
\end{align}
Consequently we have $\hb{i=1,\dots,7} = 0$ for $\ell+m=\text{odd}$ and
$\hb{i=8,9,10} = 0$ for $\ell+m=\text{even}$.

The gauge equations can be used to reduce the number of fields that need to be
solved for simultaneously. For example, for radiative modes ($\omega_m\neq0$)
in the odd sector one can solve for the
$\hb{9}$ and $\hb{10}$ fields from which the $\hb{8}$ field can be
constructed algebraically from the gauge equation \eqref{eq:gauge_odd}. Similarly,
for radiative modes in the even sector the number of field equations to be
solved simultaneously can be reduced by using
Eqs.~\eqref{eq:gauge_even1}--\eqref{eq:gauge_even3}. In this work we opt to
only use Eqs.~\eqref{eq:gauge_even2} and \eqref{eq:gauge_even3} to reduce the
number of fields to be solved from seven to five. The remaining gauge equation
\eqref{eq:gauge_even1} can then be used as a consistency check on the final
result.

The static modes ($\omega_m=0$) require a different treatment. In the odd
sector both $\hb{9}$ and $\hb{10}$ are zero as their sources vanish. The
resulting equation for $\hb{8}$ can be solved for analytically -- see Ref.~\cite{Barack-Lousto-2005} for details. In the even sector, the gauge equations \eqref{eq:gauge_even2}
and \eqref{eq:gauge_even3} can be used to eliminate the $\hb{6}$ and $\hb{7}$
fields which appear in Eqs.~\eqref{eq:eq_R1}, \eqref{eq:eq_R3} and
\eqref{eq:eq_R5}. The resulting set of three ordinary differential
equations were first solved numerically \cite{Akcay-GSF-circular}, but more
recently analytic solutions have been derived \cite{Osburn_etal:2014}. In this
work we opt for the numerical approach.

For the non-radiative low-multipole modes ($\ell=0,1,m=0$) analytic solutions are
known \cite{Detweiler-Poisson}. The $\ell=m=1$ mode is solved for numerically much
as the other radiative even sector modes are, except for this mode $\hb{7}=0$
identically. We overview this hierarchical structure for solving the field
equations in Table \ref{table:hierachical_structure}.

\begin{table}[htb!]
\centering
\begin{tabularx}{\columnwidth}{c c | l | l }
\toprule
$\ell$		& $m$ 		& $\ell+m = \text{even}$									& $\ell+m = \text{odd}$						\\	
\hline
0		& 0			& $(i) = 1,3\rightarrow6$\quad(A)							& --									\\
1		& 0			& --														& $(i) = 8$\quad(A) 					\\
1		& 1			& $(i) = 1,3,5,6\rightarrow2,4$								& --									\\
$\ge2$	& 0			& $(i) = 1,3,5\rightarrow6,7$\quad(A\textsuperscript{*})	& $(i) = 8$\quad(A)						\\
$\ge2$	& $m\neq0$	& $(i) = 1,3,5,6,7\rightarrow2,4$							& $(i) = 9,10\rightarrow 8$				\\
\toprule
\end{tabularx}
\caption{Hierarchical structure for solving the field equations. A
`$\rightarrow$' implies the field(s) to the right should be algebraically
constructed from the fields to the left using
Eqs.~\eqref{eq:gauge_even1}--\eqref{eq:gauge_odd}. An (A) implies analytic
solutions are known, and we employ them in this work except in the case of the
even static modes. }\label{table:hierachical_structure}
\end{table}

\subsection{Retarded-field solution}\label{sec:ret_solution}

In this section we outline the calculation of the retarded-field solution to
Eq.~\eqref{eq:metric_pert_FD} using the standard variation of parameters
method. In this approach the inhomogeneous solution is constructed by
multiplying the homogeneous solutions by suitable weighting coefficients. In
each sector (odd/even, static/radiative) we must solve for $k$ coupled
equations and correspondingly the space of homogeneous solutions will be $2k$
dimensional. Using $j=1,\dots,k$ as an index for the basis of homogeneous
solutions, let us define the `inner' and `outer' homogeneous solutions to the
field equation by $\hbm{i}_j$ and $\hbp{i}_j$, respectively. The inner solutions
are regular at the horizon but diverge as $r\rightarrow\infty$. Conversely, the
outer solutions are regular at spatial infinity and diverge at the horizon. For
the radiative modes the retarded solutions are selected by ensuring radiation
at the horizon is purely in-going and radiation at spatial infinity is purely
outgoing. This in turn implies that the asymptotic behaviour of the inner and outer
solutions go as
\begin{equation}
\label{eq:asymptotic_BCs}
	\hbpm{i}(r_*\to\pm\infty) \sim e^{\pm i\omega r_*},
\end{equation}
where $r_*$ is the tortoise radial coordinate defined by $dr_*/dr = f(r)^{-1}$.
A more in depth discussion of the asymptotic behavior of the radial fields is
given in Refs.~\cite{Akcay-GSF-circular,Akcay-Warburton-Barack}.

With the above definitions, the standard variation of parameters approach can
be used to construct the inhomogeneous solutions to
Eq.~\eqref{eq:metric_pert_FD} via
\begin{align}
	\hb{i}(r) = \sum_{j=1}^k\left( C_j^-(r)\hbm{i}_j(r) + C_j^+(r)\hbp{i}_j(r)\right).
\end{align}
To compute the weighting coefficients $C^\pm_j$ we define a $2k\times2k$ matrix of homogeneous solutions by
\begin{eqnarray}\label{eq:Phi_matrix}
	\arraycolsep=1.4pt\def\arraystretch{1.5}
	\Phi(r) = \left(\begin{array}{c | c}-\hbm{i}_j & \hbp{i}_j	\\ \hline -\partial_r \hbm{i}_j & \partial_r \hbp{i}_j \end{array}\right)\c
\end{eqnarray}
The weighting coefficients $C^\pm_j(r)$ are then computed with the standard variation of parameters prescription:
\begin{align}
	\left(\begin{array}{c} C^-_j(r) \\ C^+_j(r)\end{array}\right) = \int^b_a \Phi^{-1}(r')\left(\begin{array}{c} \mathbf{0} \\ \mathcal{J}^{(j)}(r')\delta(r'-r_0)\end{array}\right)\,dr',
\end{align}
where the limits on the integral depend upon which weighting coefficient is being solved for. For the $C^-_j$'s $a=r,b=\infty$ and for the $C^+_j$'s $a=2M,b=r$. The source vector is given by $k$ zeros followed by the $k$ sources from the right-hand side of the field equation \eqref{eq:metric_pert_FD}.

The delta function in the source means the integration can be done analytically and the inhomogeneous solutions can be written explicitly as
\begin{align}\label{eq:hret_construction}
	\hb{i}(r) = \left\{\begin{array}{cc} \sum_{j=1}^k C_{j0}^+ \hbp{i}_j(r)	& r\ge r_0	\\ \sum_{j=1}^k C_{j0}^- \hbm{i}_j(r)	& r\le r_0\end{array}\right.,
\end{align}
where
\begin{align}\label{eq:ret_weighting_coeffs}
	\left(\begin{array}{c} C^-_{j0} \\ C^+_{j0}\end{array}\right) =  \Phi^{-1}(r_0)\left(\begin{array}{c} \mathbf{0} \\ \mathcal{J}^{(j)}(r_0)\end{array}\right)\p
\end{align}
Note that the $C^\pm_{j0}$ are $r$-independent constants.

\section{Regularization}\label{sec:regularization}

Building on the work of Mino, Sasaki and Tanaka \cite{Mino-Sasaki-Tanaka} and
Quinn and Wald \cite{Quinn-Wald}, Detweiler and Whiting showed that the
gravitational self-force can be computed as the derivative of a suitable
regular metric perturbation, $\bar{h}^R_{\mu\nu}$, via
\begin{align}\label{eq:F_self}
	F^\mu_\self(x_0) =  \mu k^{\mu\nu\gamma\delta}\nabla_\delta\bar{h}^R_{\nu\gamma}(x_0),
\end{align}
where
\begin{align}\label{eq:k}
	k^{\mu\nu\gamma\delta} =& \half g^{\mu\delta}u^\nu u^\gamma - g^{\mu\nu}u^\gamma u^\delta - \half u^\mu u^\nu u^\gamma u^\delta \nonumber \\
								 &+ \frac{1}{4}u^\mu g^{\nu\gamma}u^\delta + \frac{1}{4}g^{\mu\delta}g^{\nu\gamma}
\end{align}
includes a projection operator that ensures the self-force is orthogonal to the
particle's four-velocity.

The regular metric perturbation is constructed by subtracting an appropriate
singular perturbation, $\bar{h}^S_{\mu\nu}$, from the usual retarded
metric perturbation $\bar{h}^{\text{ret}}_{\mu\nu}$, i.e.,
\begin{align}\label{eq:hR}
	\bar{h}^R_{\mu\nu}(x_0) = \lim_{x\rightarrow x_0}\left[\bar{h}^{\text{ret}}_{\mu\nu}(x) - \bar{h}^S_{\mu\nu}(x)\right].
\end{align}
The construction of an appropriate singular field was discussed at length in
Refs.~\cite{Detweiler-Whiting,Poisson-review}. One of the key features of the
three metric perturbations $\bar{h}^{\text{ret}/S/R}_{\mu\nu}$ is that they obey the
field equations
\begin{align}
	\mathring{\square}\bar{h}^{\text{ret}/S}_{\mu\nu} + 2 \mathring{R}^{\rho\;\;\sigma}_{\;\;\mu\;\;\nu}\bar{h}^{\text{ret}/S}_{\rho\sigma} &= -16\pi T_{\mu\nu},		\\
	\mathring{\square}\bar{h}^{R}_{\mu\nu} + 2 \mathring{R}^{\rho\;\;\sigma}_{\;\;\mu\;\;\nu}\bar{h}^{R}_{\rho\sigma} &= 0,
\end{align}
from which we see that the retarded and singular perturbations diverge in the
same way at the particle's location\footnote{Note that the retarded and singular
perturbations are a solution of the same equation, but with different boundary
conditions. As a result, it is only their local, singular behaviour which
agrees.}, whilst their difference --- the regular
perturbation --- is smooth there. Using Eqs.~\eqref{eq:F_self} and \eqref{eq:hR}
we can write the self-force as
\begin{align}
	F^\mu_{\self}(x_0) 	&= \mu \lim_{x\rightarrow x_0} \left[k^{\mu\nu\gamma\delta}\mathring{\nabla}_\delta(\bar{h}^\text{ret}_{\nu\gamma}(x) - \bar{h}^S_{\nu\gamma}(x))\right], \nonumber\\
							&= \lim_{x\rightarrow x_0} \left[F_\ret^\mu(x) - F_S^\mu\right(x)],			\label{eq:F_self2}
\end{align}
where
\begin{align}
	F_{\ret/S}^\mu(x) \equiv \mu k^{\mu\nu\gamma\delta}\mathring{\nabla}_\delta\bar{h}^{\text{ret}/S}_{\nu\gamma}(x).
\end{align}
The divergence of $\bar{h}^{\text{ret}/S}_{\mu\nu}$ at the particle makes
it challenging to work with Eq.~\eqref{eq:F_self2} directly. Consequently, a
number of reformulations have been devised to allow the gravitational
self-force to be computed. Two of these schemes, the mode-sum method and the
effective-source approach, we discuss now.

\subsection{Mode-sum method}\label{sec:mode_sum}

The key observation behind the mode-sum method is that although the full
retarded and singular metric perturbations are divergent at the particle, their
individual multipole modes remain finite everywhere. The subtraction between the
retarded and singular contributions in Eq.~\eqref{eq:F_self2} can then be made
on a mode-by-mode basis. Explicitly we can write
\begin{align}\label{eq:F_self_l}
	F_\self^\mu(x_0) = \lim_{x\rightarrow x_0}\sum_{\ellh=0}^\infty\left[F^{\mu\ellh}_{\ret}(x) - F^{\mu\ellh}_{S}(x)\right],
\end{align}
where a superscript $\ellh$ denotes a quantity's decomposition into
\textit{scalar} spherical-harmonic modes and summed over $m$, i.e.,
\begin{align}
	F^{\mu\ellh}_{\ret/S} = \sum_{m=-\ellh}^{\ellh} Y_{\ellh m}(\pi/2,\varphi_0)\int_0^{2\pi} \int_0^\pi  F^{\mu}_{\ret/S} Y^\ast_{\ellh m}(\theta,\varphi)\,d\Omega.
\end{align}
We discuss below how we interface the tensor-mode computation of the retarded
field outlined in Sec.~\ref{sec:ret_solution} and the standard mode-sum scheme
we are outlining now.

The individual multipole modes of the retarded and singular contributions to
the self-force, $F^{(\ret/S)\ellh}$ are $C^{-1}$. That is, they are finite at
the particle but, in general, their sided limits $r\rightarrow r_0^\pm$ yield
two different values, which we denote by $F^{\mu\hat{\ell}\pm}_{\ret/S}$,
respectively. For circular orbits there is no closed-form analytic solution for
$F_\ret^\mu$, and typically it is computed numerically. The singular field,
on the other hand, is amenable to an analytic treatment. The local structure of
the singular field was first analyzed by Mino \etal~\cite{Mino-Sasaki-Tanaka}
and Barack and Ori used these results to develop the mode-sum scheme shortly
thereafter \cite{mode-sum-orig}.

The scalar-harmonic mode-sum regularization formula for the redshift invariant
$h^R_{uu} \equiv h^R_{\mu\nu}u^\mu u^\nu$ \cite{Detweiler-circular}
and for the self-force are given by
\begin{align}
	h^R_{uu} 			&= \sum_{\ellh=0}^\infty\left(h^{(\ret) \ellh}_{\mu\nu}u^\mu u^\nu - H^{[0]}\right)		-D_H,			\label{eq:huu_mode_sum}		\\
	F_{\self}^\mu		&= \sum_{\ellh=0}^\infty\left(F^{\mu\ellh\pm}_{\ret} - F^{\mu\pm}_{[-1]} (2\ellh+1) - F^\mu_{[0]} \right) - D^\mu.	\label{eq:Fr_mode_sum}
\end{align}
The $\ellh$-independent $H^{[0]}, F^{\mu\pm}_{[-1]}, F^\mu_{[0]}, D_H, D_\mu$ are known as regularization parameters and their
value is known for generic geodesic orbits in Schwarzschild
\cite{Barack-Mino-Nakano-Ori-Sasaki} and Kerr spacetime \cite{Barack-Ori}. In general the coefficients of odd negative powers of $\ellh$ in the mode-sum formula are 
zero \cite{Detweiler-Messaritaki-Whiting} and in the Lorenz gauge $D^\mu=D_H=0$. For circular orbits the other nonzero
regularization parameters are given by
\begin{align}\label{eq:mode_sum_reg_params}
	H^{[0]}				&= \frac{4\mu}{\pi\sqrt{r_0^2 +\ang_0^2}}\mathcal{K},												\\
	F^{r\pm}_{[-1]} 	&= \mp \frac{\mu^2}{2 r_0^2}\left(1-\frac{3M}{r_0}\right)^{1/2},									\\
	F^r_{[0]}			&= \frac{\mu^2 r_0 \en_0^2}{\pi(\ang_0^2 + r_0^2)^{3/2}}\left[\mathcal{E} - 2\mathcal{K}\right],		
\end{align}
where $\mathcal{K} \equiv
\int^{\pi/2}_0(1-\tfrac{M}{r_0-2M}\sin^2x)^{-1/2}\,dx$ and $\mathcal{E} \equiv
\int^{\pi/2}_0(1-\tfrac{M}{r_0-2M}\sin^2x)^{1/2}dx$ are complete elliptic
integrals of the first and second kind, respectively.

The series in $\ellh$ in both Eqs.~\eqref{eq:huu_mode_sum} and
\eqref{eq:Fr_mode_sum} is truncated at $\ellh^{-1}$. This is sufficient to
regularize $h_{uu}$ and $F^r$, but the resulting sum over $\ellh$ converges
rather slowly, with each term going as $\ellh^{-2}$. It is possible to derive
higher-order regularization parameters \cite{Detweiler-Messaritaki-Whiting}; Ref.~\cite{Heffernan-Ottewill-Wardell}
provides the next two non-zero parameters that serve to increase the rate of
convergence of the mode-sum to $\ellh^{6}$. It is common practice in mode-sum
calculations to numerically fit for the yet higher-order unknown parameters to
further increase the rate of convergence of the mode-sum.

Lastly, as we mentioned above, we compute the retarded metric perturbation
within a tensor-harmonic decomposition, whereas the standard mode-sum approach
requires the retarded metric perturbation decomposed into scalar-harmonic modes
as input. Thus before regularizing we must project the tensor-harmonic modes of
the metric perturbation onto a basis of scalar harmonics. The projection equation takes the form
\begin{align}\label{eq:scalar_from_tensor}
	F^{\mu\ellh\pm}_{\ret} = \frac{\mu^2}{r_0^2} \sum_{m=-\ellh}^{\ellh} \sum_{p=-3}^{3}  Y^{\ellh m}(\pi/2,\varphi_p) \mathcal{F}^{\mu,\ellh+p,m}_{(p)\pm},
\end{align}
where the details of the $\mathcal{F}^{\mu,\ellh+p,m}_{(p)\pm}$ (but not the
self-force obtained after summing over $\ell$) depends on the
way in which the definition for the force (a quantity which is defined
\emph{on} the worldline) is extended \emph{off} the worldline to the whole
two-sphere. Barack and Sago \cite{Barack-Sago-circular} made the
computationally-convenient choice of $k^{abcd}$ being given by $u^a$ having a
constant value on the two-sphere, and the metric having its usual tensorial
value. Their expressions for the $\mathcal{F}^{\mu,\ellh+p,m}_{(p)\pm}$ are
rather cumbersome so we do not give them here; instead, their explicit form can
be found in Appendix~C of Ref.~\cite{Barack-Sago-circular}. Likewise, a similar
formula can be derived for $h_{uu}$ \cite{Warburton-thesis}.

The sum over $p$ in Eq.~\eqref{eq:scalar_from_tensor}
means that in order to compute the self-force by regularizing $\ellh_{\max}$
scalar-harmonic modes one must compute $(\ell_{\max} + 3)$ tensor-harmonic modes.
Similarly for $h_{uu}$ one must compute $(\ell_{\max} + 2)$ tensor modes. In
Sec.~\ref{sec:mode_sum_tensor} below we will recast the standard mode-sum
formula to use tensor modes rather than scalar modes, which will avoid this
projection step altogether.

\subsection{Effective-source approach}\label{sec:eff_source}

The effective-source approach is an alternative practical regularization scheme for handling
the divergence of the retarded field. Rather than first computing the retarded field and then
subtracting the singular piece as a post-processing step, as in the mode-sum scheme, one can instead work directly with an
equation for the regular field. This idea was first proposed in Refs.~\cite{Barack-Golbourn,Vega-Detweiler} and has the distinct advantage
of involving only regular quantities, making it applicable in a wider variety of scenarios than
the mode-sum scheme.

Using Eq.~\eqref{eq:hR} to rewrite $\bar{h}^\ret_{\mu\nu}$ in terms of $\bar{h}^R_{\mu\nu}$ and $\bar{h}^S_{\mu\nu}$, we can
rewrite Eq.~\eqref{eq:linearized_einstein} as
\begin{align}
	\mathring{\square}\bar{h}^R_{\mu\nu} + 2 \mathring{R}^{\rho\;\;\sigma}_{\;\;\mu\;\;\nu}\bar{h}^R_{\rho\sigma} = -16\pi T_{\mu\nu} - \mathring{\square}\bar{h}^S_{\mu\nu} - 2 \mathring{R}^{\rho\;\;\sigma}_{\;\;\mu\;\;\nu}\bar{h}^S_{\rho\sigma}
\end{align}
If $\bar{h}_{\mu\nu}^S$ is precisely the Detweiler-Whiting singular metric perturbation, then the two terms on the right hand
side of this equation cancel and $\bar{h}_{\mu\nu}^R$ becomes a homogeneous solution of the wave equation.
However, one typically does not have access to an exact expression for $\bar{h}_{\mu\nu}^S$ as the
Detweiler-Whiting singular metric perturbation is defined through a Hadamard parametrix \cite{Hadamard,Friedlander} which is not
even globally defined. Instead, the best one can typically do is a local expansion which is valid only in the
vicinity of the worldline. Let us denote such an approximation to $\bar{h}_{\mu\nu}^S$ by $\bar{h}_{\mu\nu}^P$. With the latter we will construct an \emph{effective source} that will allow us to directly compute the regular field at the worldline.

The puncture field $\bar{h}_{\mu\nu}^P$ is only valid near the worldline and so, to avoid ambiguities in the definition of the effective source, one must ensure that the puncture field goes to zero far from the particle. This is can be achieved by
multiplying $\bar{h}_{\mu\nu}^P$ by a window function, $\mathcal{W}$, with properties such that multiplying it by $\bar{h}_{\mu\nu}^P$ only modifies terms of higher order in the local expansion about the worldline than those which are explicitly given in $\bar{h}_{\mu\nu}^P$. In our particular case, it suffices to choose $\mathcal{W}$ such that
$\mathcal{W}(x_0) = 1$, $\mathcal{W'}(x_0) = 0$, $\mathcal{W''}(x_0) = 0$ and $\mathcal{W} = 0$ far
away from the worldline. The residual metric perturbation, $\bar{h}^\res_{\mu\nu}$, then obeys
\begin{align}\label{eq:box_phi_res}
	&\mathring{\square}\bar{h}^\res_{\mu\nu} + 2 \mathring{R}^{\rho\;\;\sigma}_{\;\;\mu\;\;\nu}\bar{h}^\res_{\rho\sigma} = 	S_\eff,		\nonumber \\
\end{align}
where the effective source is given by
\begin{align}
	& S_\eff \equiv -16\pi T_{\mu\nu} - \mathring{\square}(\mathcal{W}\bar{h}^P_{\mu\nu}) - 2 \mathring{R}^{\rho\;\;\sigma}_{\;\;\mu\;\;\nu}(\mathcal{W}\bar{h}^P_{\rho\sigma}).
\end{align}
This effective source is smooth and finite everywhere, except on the worldline where it has limited differentiability. The corresponding residual field has the properties
\begin{align}
	\bar{h}^\res_{\mu\nu}(x_0) &= \bar{h}_{\mu\nu}^R(x_0), \quad \nabla_\delta \bar{h}^\res_{\mu\nu}(x_0) = \nabla_\delta \bar{h}_{\mu\nu}^R(x_0)\c\nonumber \\
	\bar{h}^\res_{\mu\nu}(x) &= \bar{h}^\ret_{\mu\nu}(x) \quad \text{for} \quad x \not\in \operatorname{supp}(\mathcal{W})\p
\end{align}
As the residual metric perturbation coincides with the retarded metric perturbation far from the particle we can use the usual retarded metric perturbation boundary conditions when solving Eq.~\eqref{eq:box_phi_res}.

\section{Effective source in the frequency domain}\label{sec:Seff_in_FD}

\subsection{Construction of the puncture fields}

\subsubsection{Coordinate expansion of the singular field}

At the core of our calculation is an effective source for the field equations
which is constructed from an approximation to the Detweiler-Whiting singular
field. A suitable covariant expansion of the singular field is given by
\begin{equation}
\label{eq:hS-cov}
  \hb{\rm S}_{ab} =
    4 \mu g_{a}{}^\ab g_b{}^\bb \Big[
      \frac{1}{\epsilon} \frac{u_\ab u_\bb}{\sbar}
      + \mathcal{O}(\epsilon)\Big],
\end{equation}
where $\epsilon$ is an order-counting parameter, $\bar{s} \equiv (g_{\ab \bb} +
u_\ab u_\bb) \sigma^\ab \sigma^\bb$, $u^{\bar{a}}$ is the four-velocity and
$g_{\ab \bb}$ the background metric, with both defined as tensors on the
worldline (i.e. at the spacetime point $x_0$). We have also introduced the
bi-vector of parallel transport $g_{a}{}^\ab(x,x_0)$ and the Synge
world-function $\sigma(x,x_0)$, both of which are functions of the
worldline point $x_0$ and the point where the singular field is to be
evaluated, $x$.

This approximation is sufficient to produce a residual field which is finite on
the worldline and which gives the correct, regularized self-force. Several
higher order terms in this expansion are also known
\cite{Heffernan-Ottewill-Wardell} and can be incorporated into the calculation
in order to accelerate convergence. However, for clarity we illustrate the
approach with this simple low-order approximation and note that the methodology
does not fundamentally change at higher orders.

We now wish to use the approximation \eqref{eq:hS-cov} as a starting point to
compute the puncture fields $\bar{h}^{(i) {\rm P}}_{\ell m}$. To this end, we follow
previous regularization strategies \cite{Barack:2002mha,Barack-Ori:mode-sum-grav+EM,Detweiler-Messaritaki-Whiting,Haas-Poisson-mode-sum,Heffernan-Ottewill-Wardell,Heffernan-Ottewill-Wardell:Kerr} by introducing a Riemann normal coordinate
system in the vicinity of the worldline, and rewrite Eq.~\eqref{eq:hS-cov} as a
coordinate expansion in terms of these coordinates. Specifically, we assume
that the spacetime can be represented in terms of a spherical coordinate system
with polar and azimuthal coordinates $\alpha$ and $\beta$, radius $r$ and time
$t$. Note that, although our focus here is on the Schwarzschild spacetime, the
assumption of a spherical coordinate system does not necessarily limit us to
spherical symmetry; for example, the method works equally well in the
non-spherically symmetric Kerr spacetime \cite{Heffernan-Ottewill-Wardell:Kerr}.

Now, orienting our coordinate system such that the worldline is instantaneously
at $\alpha=0$, we define the Riemann normal coordinates $w_1 = 2 \sin
\tfrac{\alpha}{2} \cos \beta$ and $w_2 = 2 \sin \tfrac{\alpha}{2} \sin \beta$.
Using coordinate expansions of $g_{a}{}^\ab(x,x_0)$ and $\sigma(x,x_0)$
about $x=x_0$ to linear order in $x-x_0$, we obtain an approximation to
\eqref{eq:hS-cov} in terms of the $(t,r,w_1,w_2)$ Riemann normal coordinate
system. Structurally, our coordinate expansion has the form
\begin{equation} \label{eq:hS-coord}
  \hb{\rm S}_{ab} =
    \frac{1}{\epsilon}\frac{c^{(1)}_{ab}}{\rho}
    + \epsilon^0 \Big[
      \frac{c^{(2)}_{ab} \Delta r}{\rho}
      + \frac{c^{(3)}_{ab} \Delta r^3}{\rho^3} \Big]
    + \mathcal{O}(\epsilon),
\end{equation}
where $\rho$ is the leading-order term in the coordinate expansion of $\sbar$
and the coefficients $c^{(1)}_{ab}$, $c^{(2)}_{ab}$ and $c^{(3)}_{ab}$ do not
depend on $\Delta r$ or $\alpha$ (and hence $w_1$ and $w_2$)\footnote{This form is valid
for the case of circular orbits in Schwarzschild spacetime, where any quadratic
dependence on $w_1$ and $w_2$ can be replaced with a term involving $\rho^2$ and
$\Delta r^2$. The structure is slightly more complicated in more general cases
where odd powers of $w_1$ and $w_2$ can appear, but nonetheless the following
analysis remains qualitatively unchanged.}.
The coefficients are also independent of $t$ since we
have chosen $\Delta t = 0$, i.e., $x$ and $x_0$ are points on the same time
slice. There is still a potential time dependence, however, through the
dependence of the coefficients on the worldline and four-velocity.

In the next subsection, we will seek a decomposition into spherical-harmonic
modes. We therefore apply the (approximate) Jacobian from $(w_1,w_2)$
coordinates to $(\alpha,\beta)$ coordinates. In doing so, we pull out a factor
of $\sin \alpha$ from the Jacobian when computing $h_{t\beta}$, $h_{r\beta}$
and $h_{\alpha\beta}$, and a factor of $\sin^2 \alpha$ when computing
$h_{\beta\beta}$. The reason for doing so will become clear during the mode
decomposition, and is related to the fact that the Riemann normal coordinate
system is regular on the worldline, but the $(\alpha,\beta)$ coordinate system
is not.

Evaluating Eq.~\eqref{eq:hS-cov} for our particular case of a circular orbit in
Schwarzschild spacetime, we arrive at our desired coordinate expansion of the
Detweiler-Whiting singular metric perturbation. With Riemann normal components
given by
\begin{subequations}
\begin{align}
  \bar{h}_{tw_1} &= -\frac{1}{\rho} \Big[
    \frac{4 r_0^2 \Omega_\varphi (r_0-2 M)}{r_0-3 M}
    +\frac{2 \Delta r\, r_0 \Omega_\varphi}{r_0-3 M} \times
  \nonumber \\
    & \qquad \frac{r_0^2-3 M r_0+2 M^2-2 M^2 \sin^2 \beta}{(r_0 - 2M)(1-\tfrac{M}{r_0-2M} \sin^2 \beta)}
  \Big],
\\
  \bar{h}_{r w_1} &= \frac{4M r_0 \sin \alpha\,\cos \beta}{\rho(r_0-3 M)},
\\
  \bar{h}_{w_1 w_1} &= \frac{\cos^2\beta}{\rho} \Big[
    \frac{4 M r_0^2}{r_0-3 M}
    +\frac{2 \Delta r\, M r_0}{r_0-3 M} \times
  \nonumber \\
    & \qquad \frac{3 r_0 - 7 M - 2 M \sin^2 \beta}{(r_0 - 2M)(1-\tfrac{M}{r_0-2M} \sin^2 \beta)}
  \Big].
\end{align}
\end{subequations}
our approximation to the Detweiler-Whiting singular metric is then given by
\begin{subequations}
\begin{align}
  \bar{h}_{tt} &= \frac{1}{\rho} \Big[
    \frac{4 (r_0-2 M)^2}{r_0 (r_0-3M)}
    - \frac{2 \Delta r}{r_0^2 (r_0-3M)} \times
  \nonumber \\
    & \quad\frac{r_0^2-7 M r_0+10 M^2-2 M (r_0-4 M)\sin^2 \beta}{1-\tfrac{M}{r_0-2M} \sin^2 \beta}
    \Big]
\\
  \bar{h}_{tr} &= -\frac{4 r_0 \Omega_\varphi (r_0-2 M) \sin \alpha \cos \beta}{\rho (r_0-3 M)}
\\
  \bar{h}_{t\alpha} &= \bar{h}_{t w_1} \cos \beta,
\\
  \bar{h}_{t\beta} &= - \bar{h}_{t w_1} \sin \alpha \sin \beta,
\\
  \bar{h}_{rr} &= 0,
\\
  \bar{h}_{r\alpha} &= \bar{h}_{r w_1} \cos \beta,
\\
  \bar{h}_{r\beta} &= - \bar{h}_{r w_1} \sin \alpha \sin \beta,
\\
  \bar{h}_{\alpha\alpha} &= \bar{h}_{w_1 w_1} \cos^2 \beta,
\\
  \bar{h}_{\alpha\beta} &= - \bar{h}_{w_1 w_1} \sin \alpha \sin \beta \cos \beta,
\\
  \bar{h}_{\beta\beta} &= \bar{h}_{w_1 w_1} \sin^2 \alpha \sin^2 \beta.
\end{align}
\end{subequations}
This approximation includes all contributions at order $\epsilon^{-1}$ and
$\epsilon^0$, with the exception of terms proportional to $\Delta r^3 / \rho^3$,
which we neglect as their mode decomposition yields only terms proportional to
$\Delta r^2$ and higher.

\subsubsection{Mode decomposition}

We now proceed with the decomposition of our coordinate expansion into tensor
spherical-harmonic modes. For this, we must evaluate the integrals of the
singular field against the tensor spherical harmonics,
\begin{equation}
  \label{eq:hPilm-tr}
  \bar{h}^{(i)P}_{\ell m} = \frac{r}{\mu\, a^{(i)}_\ell} \int_{0}^{2 \pi} \int_{0}^{\pi} \bar{h}_{\tau\kappa} \eta^{\tau\mu}\eta^{\kappa\nu} Y_{\mu\nu}^{(i)\ell m}{}^\ast \sin \alpha \, d\alpha \, d\beta.
\end{equation}
For the circular orbit case we are considering here, the explicit form for the
integrand for each $i={1,\ldots,10}$ field is given in
Table~\ref{table:puncture-integrands}.
\begin{table*}[htb!]
\begin{center}
\begin{tabular}{c|c}
\toprule
$(i,\ell,m')$ & Integrand \\
\hline
$(1,\ell,0)$ & $r \sqrt{\frac{2 \ell+1}{4\pi}} (\bar{h}_{tt} + f^2 \bar{h}_{rr}) P_\ell^0(\cos \alpha) \sin \alpha$ \\
\hline \\
$(2,\ell,\pm1)$ & $ \pm 2 r \sqrt{\frac{2 \ell+1}{4\pi \ell(\ell+1)}} f \bar{h}_{tr} \cos \beta P_\ell^1(\cos \alpha) \sin \alpha$ \\
\hline \\
$(3,\ell,0)$ & $r \sqrt{\frac{2 \ell+1}{4\pi}} f (\bar{h}_{tt} - f^2 \bar{h}_{rr}) P_\ell^0(\cos \alpha) \sin \alpha$ \\
\hline \\
$(4,\ell,\pm1)$ & $\pm 2 \sqrt{\frac{(2 \ell+1)\ell(\ell+1)}{4\pi}} \frac{\bar{h}_{tw_1}}{\ell(\ell+1)}
  \Big[ \sin^2\beta P_\ell^1(\cos \alpha)
        + \cos^2 \beta\frac{\ell^2 P_{\ell+1}^1(\cos \alpha)-(\ell+1)^2 P_{\ell-1}^1(\cos \alpha)}{2 \ell + 1}\Big]$ \\
\hline \\
$(5,\ell,0)$ & $2 \sqrt{\frac{2 \ell+1}{4\pi}} f \bar{h}_{r w_1} \cos\beta \frac{\ell (\ell+1)}{2\ell+1}\Big[P_{\ell+1}^0(\cos \alpha) - P_{\ell-1}^0(\cos \alpha)\Big] $ \\
\hline \\
$(5,\ell,\pm2)$ & $2 \sqrt{\frac{(2 \ell+1)\ell(\ell+1)}{4\pi (\ell-1)(\ell+2)}} f \bar{h}_{r w_1} \cos\beta \frac{1}{\ell (\ell+1)}\Big[4 \sin^2\beta P_\ell^2(\cos \alpha) + (\cos^2\beta - \sin^2\beta) \frac{(\ell-1) \ell P_{\ell+1}^2(\cos \alpha) - (\ell+1) (\ell+2) P_{\ell-1}^2(\cos \alpha)}{2 \ell + 1}\Big]$ \\
\hline \\
$(6,\ell,0)$ & $\frac{1}{r} \sqrt{\frac{2 \ell+1}{4\pi}} \bar{h}_{w_1 w_1} P_\ell^0(\cos \alpha) \sin \alpha$ \\
\hline \\
$(7,\ell,\pm2)$ & $\begin{aligned}\tfrac{2}{r} &\sqrt{\tfrac{(2 \ell+1)(\ell-1)\ell(\ell+1)(\ell+2)}{4\pi}} \tfrac{1}{(\ell-1)\ell(\ell+1)(\ell+2)} \tfrac{\bar{h}_{w_1 w_1}}{\sin\alpha} \Big[8 \cos^2\beta \sin^2\beta \tfrac{(\ell-1)^2 P_{\ell+1}^2(\cos\alpha) - (\ell+2)^2 P_{\ell-1}^2(\cos\alpha)}{2\ell+1} \\
& + (\cos^2\beta - \sin^2\beta)^2 \tfrac{(\ell-1)^2 \ell^2 (2\ell-1) P_{\ell+2}^{2}(\cos\alpha) - 2 (\ell-3) (\ell-1) (\ell+2) (\ell+4) (2\ell+1) P_\ell^2(\cos\alpha) + (\ell+1)^2 (\ell+2)^2 (2\ell+3) P_{\ell-2}^2(\cos\alpha)}{2 (2\ell-1) (2 \ell + 1) (2\ell+3)}\Big]\end{aligned}$ \\
\hline \\
$(8,\ell,\pm1)$ & $-2 \sqrt{\frac{(2 \ell+1)\ell(\ell+1)}{4\pi}} \frac{i \, \bar{h}_{tw_1}}{\ell(\ell+1)}
  \Big[ P_\ell^1(\cos \alpha) \cos^2\beta
        +\frac{\ell^2 P_{\ell+1}^1(\cos \alpha)-(\ell+1)^2 P_{\ell-1}^1(\cos \alpha)}{2 \ell + 1} \sin^2 \beta\Big]$ \\
\hline \\
$(9,\ell,\pm2)$ & $\mp \sqrt{\frac{(2 \ell+1)\ell(\ell+1)}{4\pi (\ell-1)(\ell+2)}} 4 i f \bar{h}_{r w_1} \cos\beta \frac{1}{\ell (\ell+1)} \Big[(\cos^2\beta - \sin^2\beta) P_\ell^2(\cos \alpha) + \sin^2\beta \frac{(\ell-1) \ell P_{\ell+1}^2(\cos \alpha) - (\ell+1) (\ell+2) P_{\ell-1}^2(\cos \alpha)}{2 \ell + 1}\Big]$ \\
\hline \\
$(10,\ell,\pm2)$ & $\begin{aligned}\mp\tfrac{1}{r} &\sqrt{\tfrac{(2 \ell+1)(\ell-1)\ell(\ell+1)(\ell+2)}{4\pi}} \tfrac{1}{(\ell-1)\ell(\ell+1)(\ell+2)} \tfrac{4 i \bar{h}_{w_1 w_1}}{\sin\alpha} \Big[(\cos^2\beta - \sin^2\beta)^2 \tfrac{(\ell-1)^2 P_{\ell+1}^2(\cos\alpha) - (\ell+2)^2 P_{\ell-1}^2(\cos\alpha)}{2\ell+1} \\
& + \cos^2\beta \sin^2\beta \tfrac{(\ell-1)^2 \ell^2 (2\ell-1) P_{\ell+2}^{2}(\cos\alpha) - 2 (\ell-3) (\ell-1) (\ell+2) (\ell+4) (2\ell+1) P_\ell^2(\cos\alpha) + (\ell+1)^2 (\ell+2)^2 (2\ell+3) P_{\ell-2}^2(\cos\alpha)}{(2\ell-1) (2 \ell + 1) (2\ell+3)}\Big]\end{aligned}$ \\
\toprule
\end{tabular}
\caption{
Integrands appearing in the mode decomposition of all ten
tensor-harmonic components of the singular metric perturbation for the case
of a circular geodesic orbit in Schwarzschild spacetime.}\label{table:puncture-integrands}
\end{center}
\end{table*}

\begin{figure}[htb!]
  \includegraphics[width=8.5cm]{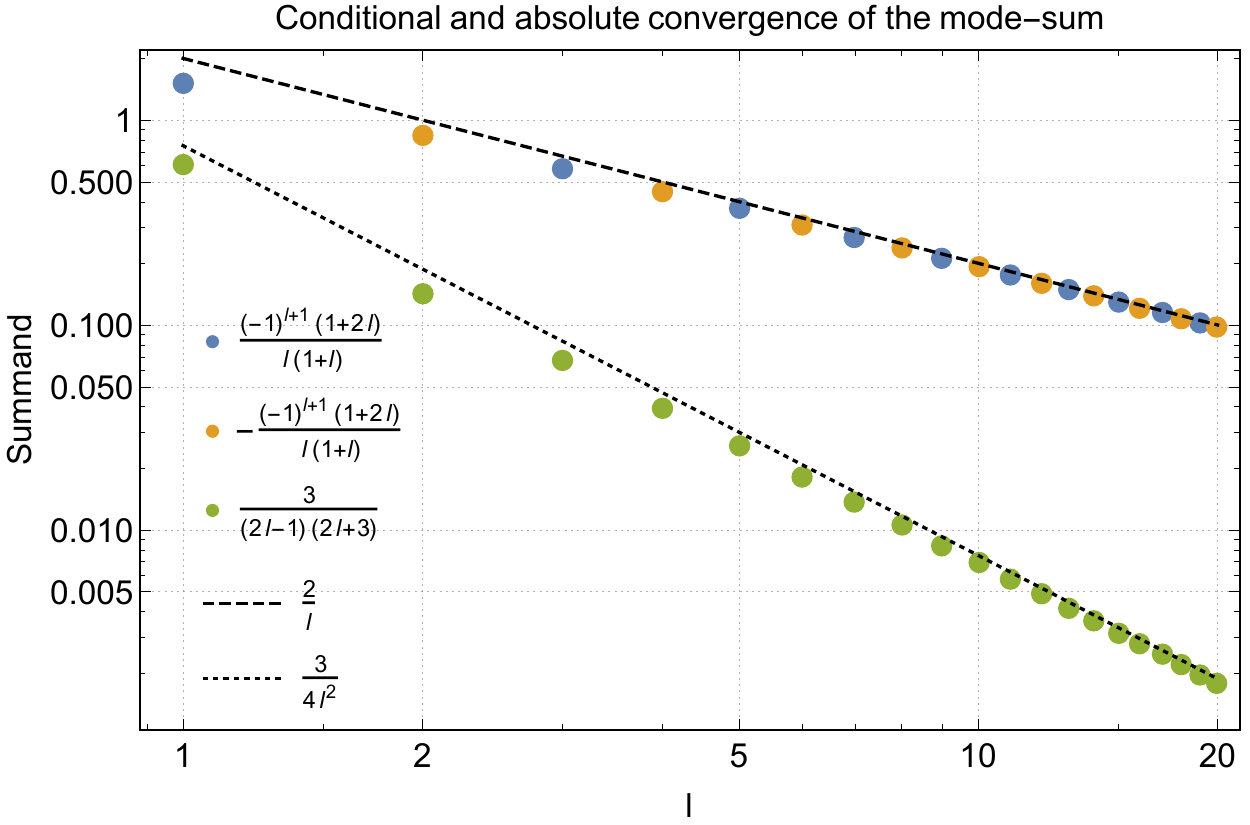}
  \caption{Effect of a spurious non-smoothness away from the worldline on the
           convergence of a mode-sum scheme near the worldline. The
           non-smoothness manifests itself as a term of the form
           $\tfrac{(-1)^{\ell+1}(2\ell+1)}{\ell(\ell+1)}$, and appears to spoil
           any hope of rapid convergence (blue/orange dots). This can be
           mitigated by using a smoothing factor which converts the
           conditionally convergent behaviour into a more rapid
           absolutely convergent behaviour (green). The final result is not
           altered since the infinite sum of conditionally convergent terms
           is exactly equal to the infinite sum of absolutely convergent terms.}
\label{fig:spurious-mode-sum}
\end{figure}

The mode decomposition works much the same as with the scalar-field
case described in Ref.~\cite{Warburton-Wardell}. There are, however, some key
differences which introduce additional complexity to the gravitational case:
\begin{enumerate}
  \item The fact that we have tensor (as opposed to scalar) harmonics makes
        the mode decomposition integrals slightly more involved.
  \item Whereas in the scalar case we only required the $m'=0$ modes, we now
        require $m'=0$ for $\bar{h}^{(1)}_{\ell m}$, $\bar{h}^{(3)}_{\ell m}$ and
        $\bar{h}^{(6)}_{\ell m}$, $m'=1$ for $\bar{h}^{(2)}_{\ell m}$, $\bar{h}^{(4)}_{\ell m}$
        and $\bar{h}^{(8)}_{\ell m}$, $m'=0,2$ for $\bar{h}^{(5)}_{\ell m}$, and $m'=2$ for 
        $\bar{h}^{(7)}_{\ell m}$, $\bar{h}^{(9)}_{\ell m}$ and $\bar{h}^{(10)}_{\ell m}$. This is
        because we would like to compute the metric perturbation and its
        derivative (for the self-force) on the worldline, and these are the 
        only modes which do not vanish on the worldline, at $\alpha=0$. Note that
        in principle other modes could contribute ($m'=1$ for
        $\bar{h}^{(1)}_{\ell m}$, $\bar{h}^{(3)}_{\ell m}$ and $\bar{h}^{(6)}_{\ell m}$,
        $m'=0$ for $\bar{h}^{(2)}_{\ell m}$, $m'=0,2$ for $\bar{h}^{(4)}_{\ell m}$
        and $\bar{h}^{(8)}_{\ell m}$, $m'=1$ for $\bar{h}^{(5)}_{\ell m}$ and
        $\bar{h}^{(9)}_{\ell m}$, and $m'=1,3$ for $\bar{h}^{(7)}_{\ell m}$ and
        $\bar{h}^{(10)}_{\ell m}$), but the integrals for those modes all contain odd
        powers of $\sin \beta$ or $\cos \beta$ and therefore their contribution
        vanishes after integration over $\beta$.
  \item The coordinate approximation we are using for the singular field has a
        spurious non-smoothness away from the worldline, at $\alpha=\pi$. This
        can be seen in $\rho$, which has a $\beta$-direction dependent limit
        as $\alpha \to \pi$. This problem did not manifest itself in the scalar
        case, since the isotropic nature of the $m'=0$ mode means it cannot
        include any information about direction dependence.
\end{enumerate}
The first two items above do not cause any fundamental issues, they merely
add some extra algebraic complexity to the problem. The third item, however,
does cause problems if not handled appropriately. The non-smoothness introduces
a spurious component in the puncture which behaves as $\tfrac{(-1)^\ell}{\ell}$
in a mode-sum formula such as Eqs.~\eqref{eq:huu_mode_sum} and
\eqref{eq:Fr_mode_sum}. This renders the sum not absolutely convergent, although
the $(-1)^\ell$ factor means that it is in fact conditionally convergent since,
for example, 
$\sum_{\ell=1}^{\infty} \tfrac{(-1)^{\ell+1}(2\ell+1)}{\ell(\ell+1)} = 1$.
In practice, this makes the sum over modes converge very slowly, see
Fig.~\ref{fig:spurious-mode-sum}.

Fortunately, there is a straightforward resolution to this problem. A smooth
window function, $\win_{m'}(\alpha)$, in the $\alpha$ direction is effective in
eliminating the spurious non-smoothness affecting the modes. To ensure the
self-force is not affected, we require that $\win_{m'}(\alpha) \sim 1 + \mathcal{O}(\alpha^2)$
near $\alpha=0$, while eliminating the effect of the non-smoothness on a
particular $m'$ mode requires $\win_{m'}(\alpha) \sim (\pi-\alpha)^{\lceil m'/2
\rceil}$ near $\alpha=\pi$, where $\lceil m'/2 \rceil$ is the smallest integer
greater than or equal to $m'/2$. We make the particular choice $\win_{m'}(\alpha) =
(\cos \tfrac{\alpha}{2})^{\lceil m'/2 \rceil}$, which satisfies both of the
above criteria.

\subsubsection{Integrals over \texorpdfstring{$\alpha$}{alpha}}

In the circular geodesic case, the quantity $\rho$ appearing in the singular
metric perturbation is given by
\begin{equation}
  \rho^2 = \frac{2 \chi r_0^2(r_0-2M)}{r_0-3M} (\delta^2 + 1 - \cos \alpha),
\end{equation}
where
\begin{equation}
  \delta^2 \equiv \frac{\Delta r^2}{2\chi r_0}\frac{r_0-3M}{(r_0-2M)^2} ,
\end{equation}
and
\begin{equation}
  \chi \equiv 1-\frac{M}{r_0-2M} \sin^2 \beta.
\end{equation}
Then, the integrals over $\alpha$ all take one of nine possible forms which can
be evaluated analytically. In our particular case, we are only interested in
the behaviour at the leading two orders in $\Delta r \equiv r-r_0$.
To simplify our expressions, we introduce
\begin{align}
	\Lambda_1 \equiv \Lambda_{1,0} = \tfrac{\ell (\ell+1)}{(2\ell-1) (2\ell+3)},		\label{eq:Lambda_1}
\end{align} and
\begin{align}
	\Lambda_2 \equiv \Lambda_{2,0} = \tfrac{(\ell-1)\ell(\ell+1)(\ell+2)}{(2\ell-3) (2\ell-1) (2\ell+3) (2\ell+5)},		\label{eq:Lambda_2}
\end{align}
where $\Lambda_{m,n}$ is defined later in Eq.~\eqref{eq:tensor-rp-ldep}.
Then, for $\bar{h}^{(1)}_{\ell0}$, $\bar{h}^{(3)}_{\ell0}$ and
$\bar{h}^{(6)}_{\ell0}$ the $\Delta r$-expanded integrals are given by
\begin{eqnarray}
  \int_0^\pi && \frac{P_\ell^0(\cos \alpha) \sin \alpha}{(\delta^2 + 1 - \cos \alpha)^{1/2}} \, d\alpha
  \nonumber \\
  && = \frac{1}{2\ell+1}\Big[ 2 \sqrt{2} - 2 (2\ell+1)|\delta| + \mathcal{O}(\delta^2) \Big].
\end{eqnarray}
For $\bar{h}^{(2)}_{\ell1}$, they are given by
\begin{eqnarray}
  \int_0^\pi && \frac{P_\ell^1(\cos \alpha) \sin^2 \alpha}{(\delta^2 + 1 - \cos \alpha)^{1/2}} \, d\alpha
  \nonumber \\
  && = \frac{1}{2\ell+1}\Big[-8 \sqrt{2} \Lambda_1 + \mathcal{O}(\delta^2) \Big],
\end{eqnarray}
For $\bar{h}^{(4)}_{\ell1}$ and $\bar{h}^{(8)}_{\ell1}$ they are given by
\begin{eqnarray}
  \int_0^\pi && \frac{\win_1(\alpha)}{\ell(\ell+1)} \frac{\ell^2 P_{\ell+1}^1(\cos \alpha)-(\ell+1)^2 P_{\ell-1}^1(\cos \alpha)}{(2\ell+1)(\delta^2 + 1 - \cos \alpha)^{1/2}} \, d\alpha
  \nonumber \\
  && = \frac{1}{2\ell+1}\Big[-\frac{6 \sqrt{2}}{(2\ell-1) (2\ell+3)} + (2\ell+1) |\delta| + \mathcal{O}(\delta^2) \Big],
  \nonumber \\
\end{eqnarray}
and
\begin{eqnarray}
  \int_0^\pi && \frac{\win_1(\alpha)}{\ell(\ell+1)} \frac{P_{\ell}^1(\cos \alpha)}{(\delta^2 + 1 - \cos \alpha)^{1/2}} \, d\alpha
  \nonumber \\
  && = \frac{1}{2\ell+1}\Big[-8 \sqrt{2} \Lambda_1 + \frac{6 \sqrt{2}}{(2\ell-1) (2\ell+3)} \nonumber \\
  && \qquad \qquad \qquad + (2\ell+1)|\delta| + \mathcal{O}(\delta^2) \Big].
\end{eqnarray}
For $\bar{h}^{(5)}_{\ell0}$ they are given by
\begin{eqnarray}
  \int_0^\pi && \frac{\ell(\ell+1)}{(2\ell+1)}\frac{\big[P_{\ell+1}^0(\cos \alpha) - P_{\ell-1}^0(\cos \alpha)\big]\sin\alpha}{(\delta^2 + 1 - \cos \alpha)^{1/2}} \, d\alpha
  \nonumber \\
  && = \frac{1}{2\ell+1}\Big[-8 \sqrt{2} \Lambda_1 + \mathcal{O}(\delta^2) \Big].
\end{eqnarray}
For $\bar{h}^{(5)}_{\ell2}$ and $\bar{h}^{(9)}_{\ell2}$ they are given by
\begin{eqnarray}
  \int_0^\pi && \frac{1}{\ell(\ell+1)}\frac{\win_2(\alpha) P_\ell^2(\cos \alpha)\sin\alpha}{(\delta^2 + 1 - \cos \alpha)^{1/2}} \, d\alpha
  \nonumber \\
  && = \frac{1}{2\ell+1} \Big[ 32 \sqrt{2}\, \Lambda_2
  \nonumber \\
  && \qquad - \frac{120 \sqrt{2} (\ell-1)(\ell+2)}{(2\ell-3)(2\ell-1)(2\ell+3)(2\ell+5)} + \mathcal{O}(\delta^2) \Big],
  \nonumber \\
\end{eqnarray}
and
\begin{eqnarray}
  \int_0^\pi && \frac{\win_2(\alpha) \sin\alpha}{(2 \ell + 1)\ell(\ell+1)(\delta^2 + 1 - \cos \alpha)^{1/2}} \times
  \nonumber \\
  && \Big[(\ell-1) \ell P_{\ell+1}^2(\cos \alpha) - (\ell+1) (\ell+2) P_{\ell-1}^2(\cos \alpha)\Big] \, d\alpha
  \nonumber \\
  && = \frac{1}{2\ell+1} \Big[ -32 \sqrt{2} \Lambda_2
  \nonumber \\
  && \quad + \frac{240 \sqrt{2} (\ell-1)(\ell+2)}{(2\ell-3)(2\ell-1)(2\ell+3)(2\ell+5)}
    + \mathcal{O}(\delta^2) \Big].
\end{eqnarray}
Finally, for $\bar{h}^{(7)}_{\ell2}$ and $\bar{h}^{(10)}_{\ell2}$ they are given by
\begin{eqnarray}
  \int_0^\pi && \frac{\win_2(\alpha) \csc\alpha}{ (\ell-1)\ell(\ell+1)(\ell+2)(2\ell+1)(\delta^2 + 1 - \cos \alpha)^{1/2}} \times
  \nonumber \\
  && \Big[(\ell-1)^2 P_{\ell+1}^2(\cos\alpha) - (\ell+2)^2 P_{\ell-1}^2(\cos\alpha)\Big]\, d\alpha
  \nonumber \\
  && = \frac{1}{2\ell+1} \Big[ \frac{10 \sqrt{2}}{(2\ell-1)(2\ell+3)} - \frac{1}{4}(2\ell+1) |\delta| + \mathcal{O}(\delta^2) \Big],
  \nonumber \\
\end{eqnarray}
and
\begin{eqnarray}
  \int_0^\pi && \frac{1}{ (\ell-1)\ell(\ell+1)(\ell+2)} \times
  \nonumber \\
  && \quad  \frac{\win_2(\alpha) \csc\alpha}{(\delta^2 + 1 - \cos \alpha)^{1/2}} \frac{1}{(2\ell-1)(2\ell+1)(2\ell+3)} \times
  \nonumber \\
  &&  \Big[(\ell-1)^2 \ell^2 (2\ell-1) P_{\ell+2}^{2}(\cos\alpha)
  \nonumber \\
  &&  \quad - 2 (\ell-3) (\ell-1) (\ell+2) (\ell+4) (2\ell+1) P_\ell^2(\cos\alpha)
  \nonumber \\
  &&  \quad + (\ell+1)^2 (\ell+2)^2 (2\ell+3) P_{\ell-2}^2(\cos\alpha)\Big] \, d\alpha
  \nonumber \\
  && = \frac{1}{2\ell+1} \Big[ - 32 \sqrt{2} \, \Lambda_2 + \frac{40 \sqrt{2}}{(2\ell-1)(2\ell+3)}
  \nonumber \\
  && \qquad \qquad \qquad + (2\ell+1) |\delta| + \mathcal{O}(\delta^2) \Big].
\end{eqnarray}

\subsubsection{Integrals over \texorpdfstring{$\beta$}{beta}}

With the integrals over $\alpha$ having been evaluated analytically as a power
series in $\delta$, we are next faced with the integrals over $\beta$. The
functional dependence on $\beta$ can be rewritten in terms of integer and
half-integer powers of $\chi = 1-\tfrac{M}{r_0-2M}\sin^2\beta$. These integrals
are straightforward to evaluate and yield either polynomials in
$\tfrac{M}{r_0-2M}$, or complete elliptic integrals with argument $\tfrac{M}{r_0-2M}$. Specifically,
\begin{equation}
  \int_0^{2\pi} \chi^{n} d \beta
   = 2\pi \, {}_2 F_1 \big(n, \tfrac{1}{2}, 1, \tfrac{M}{r_0-2M} \big),
\end{equation}
which has three special cases: for $n=-1/2$ it reduces to the elliptic integral
of the first kind, $\mathcal{K}\big(\tfrac{M}{r_0-2M}\big)$; for $n=1/2$ it
reduces to the elliptic integral of the second kind,
$\mathcal{E}\big(\tfrac{M}{r_0-2M}\big)$; for $n$ an integer it is a polynomial
in $\tfrac{M}{r_0-2M}$. All other cases can be related to these three using the
recursion relation for the hypergeometric function,
\begin{equation}
\mathcal{F}_{p+1} (k) = \frac{p-1}{p \left(k - 1\right)} \mathcal{F}_{p-1}(k) + \frac{1 - 2p + \left(p - \frac{1}{2}\right) k}{p \left(k - 1\right)} \mathcal{F}_p(k),
\end{equation}
where $\mathcal{F}_p(k) \equiv {}_2 F_1 \big(p, \tfrac{1}{2}, 1, k \big)$.

\subsection{Construction of the effective source and residual fields}
\label{sec:construct_hres}

In order to construct an effective source we must choose a window function, $\win$, to confine the definition of the puncture to a neighbourhood of the worldline. As discussed in Sec.~\ref{sec:eff_source} the constraints on the window function are that $\win(x_0) = 1, \win'(x_0)=0, \win''(x_0) = 0$ and $\win=0$ far away from the worldline. These conditions leave considerable freedom when choosing a window function. In this work we shall use the window function given by
\begin{align}
	\win(r) = e^{-8M^{-4}(r-r_0)^4}.
\end{align}
We make this choice as it is easy to implement and, although not formally compact, it is effectively compact within our numerical scheme. Other authors have made different choices. Vega \etal~\cite{Vega_etal:SF_in_3+1} used a compact window function that allowed for a smooth transition from the residual to the retarded field. Alternatively, a compact source can be achieved using the worldtube approach of Barack and Golbourn \cite{Barack-Golbourn}. In Ref.~\cite{Warburton-Wardell} we used a Heaviside $\Pi$ function and showed that this was equivalent to the worldtube method. In this work we opt not to do this for ease of implementation, though we note, by building on a draft of this work, that a worldtube method has been implemented for the gravitational case \cite{Barack-Miller-Pound:priv_comm}.

With the window function chosen the effective sources are given by
\begin{align}
	S_{\ell m}^{(i)\eff} 	=& \mathcal{J}_{\ell m}^{(i)}\delta(r-r_0) - \square_{\ell m}^{sc}\left(\win\bar{h}^{(i)P}_{\ell m}\right) 	\nonumber	\\
						 &+ 4f^{-2} \mathcal{M}^{(i)}{}_{(j)}\left(\win \bar{h}^{(j)P}_{\ell m}\right).
\end{align}
For brevity we will not display the explicit form of the $S_{\ell
m}^{(i)\eff}$. Using the field equations in Appendix
\ref{apdx:field_eqs_coupling} and punctures in Appendix \ref{apdx:punctures} it
is straightforward to compute the effective sources using computer algebra
packages. However, we do point out one potential subtlety: in the above equation
we have implicitly assumed that the wave operator commutes with the mode
decomposition, an assumption which is not necessarily true. Indeed Barack and Ori
\cite{Barack:2002mha} pointed out that the mode decomposition does \emph{not}
always commute with radial derivatives; likewise from the Wigner–Eckart theorem
one may be concerned that a spherical-harmonic mode decomposition which fails
to include all modes would not commute with the angular derivatives. In the
current context both concerns turn out to be unfounded. The Barack-Ori
observation is only an issue if the limit $\Delta r \to 0$ is taken, but we
avoid doing so while computing the puncture fields. The higher
spherical-harmonic modes of the puncture that we neglect would indeed contribute to the effective
source one obtains, but only in a way which affects the higher derivatives of the
residual field (since those higher modes vanish when evaluated at $\alpha=0$).

The construction of the residual metric perturbation now proceeds as follows. Via the variation of parameters prescription we have
\begin{align}\label{eq:hres}
	\hbres{i}(r) = \sum_{j=1}^k\left( C_j^{-\text{res}}(r)\hbm{i}_j(r) + C_j^{+\text{res}}(r)\hbp{i}_j(r)\right),
\end{align}
where recall that we use $j$ to index the $k$ basis of a given $\ell m$ mode. The weighting coefficients are given by
\begin{align}\label{eq:Cres}
	\left(\begin{array}{c} C^{-\res}_j(r) \\ C^{+\res}_j(r)\end{array}\right) = \int^b_a \Phi^{-1}(r')\left(\begin{array}{c} \mathbf{0} \\ S^{(i)\eff}\end{array}\right)\,dr',
\end{align}
where $\Phi$ is the $2k\times 2k$ matrix of homogeneous solutions, defined in Eq.~\eqref{eq:Phi_matrix}. The source vector is formed of $k$ zeros followed by the $k$ effective sources. The integration limits in Eq.~\eqref{eq:Cres} depend upon which weighting coefficient is being solved for. For the $C^{-\res}_j$'s $a=r,b=\infty$ and for the $C^{+\res}_j$'s $a=2M,b=r$.

In order to compute the self-force we also require the first radial derivatives of the metric perturbation fields. These are easily constructed via
\begin{align}\label{eq:dhres}
	\hbres{i}{}'(r) = \sum_{j=1}^k\left( C_j^{-\text{res}}(r)\hbm{i}_j{}'(r) + C_j^{+\text{res}}(r)\hbp{i}_j{}'(r)\right).
\end{align}

Lastly we discuss how to construct the remaining fields using the gauge equations using the hierarchical scheme outlined in Table \ref{table:hierachical_structure}. This is achieved by noting that the gauge equations \eqref{eq:gauge_even1}-\eqref{eq:gauge_odd} are for the retarded field. On the worldline we can write $\hb{i} = \bar{h}^{(i)\res} + \bar{h}^{(i)P}$. The remaining residual fields can be obtained by substituting this split into the gauge equations and rearranging for the $\bar{h}^{(i)\res}$.

\section{Numerical implementation and results}\label{sec:num_implementation_and_results}

The self-force experienced by a particle moving along a fixed geodesic of the background Schwarzschild spacetime was first calculated in the Lorenz gauge by Barack and Sago \cite{Barack-Sago-circular}. In calculating the retarded field they used a time-domain implementation for the $\ell\ge2$ modes and used a frequency-domain method to calculate the monopole ($\ell=0$) and dipole ($\ell=1$) modes \cite{Barack-Lousto-2005,Detweiler-Poisson}. They constructed the self-force by projecting the tensor-harmonic modes of the retarded field onto a basis of scalar harmonics and regularizing using the standard mode-sum scheme. Lorenz-gauge calculations were later extended to generic bound orbits in Schwarzschild spacetime \cite{Barack-Ori-Sago,Barack-Sago-eccentric,Golbourn-thesis,Akcay-GSF-circular,Akcay-Warburton-Barack,Osburn_etal:2014}.

In this section we detail how to compute, in the frequency domain, the Lorenz-gauge self-force along a circular geodesic using the effective-source method we have developed above. Before giving the algorithm for the computation we briefly discuss how we construct numerical boundary conditions in order to solve for the retarded homogeneous metric perturbation.

\subsection{Numerical boundary conditions}\label{sec:num_BCs}

For the radiative modes ($\omega\neq0$) the asymptotic boundary conditions for the retarded-field solutions are given by Eq.~\eqref{eq:asymptotic_BCs}. In practice we cannot place the boundaries of our numerical domain at $r_* = \pm\infty$. Instead we construct boundary conditions at a finite radius by expanding the asymptotic boundary conditions in an appropriate series. For the radiative modes we use the expansions:
\begin{align}
	\hbm{i}(r_\text{in}) 	&= e^{-i\omega_m r_*^\text{in}}\sum_{k=0}^{k^-_\text{max}} b^i_k(r_\text{in}-2M)^k	,			\\
	\hbp{i}(r_\text{out}) 	&= e^{i\omega_m r_*^\text{out}}\sum_{k=0}^{k^+_\text{max}} \frac{a^i_k}{r_\text{out}^k}	,
\end{align}
where $r_*^\text{in/out} \equiv r_*(r_\text{in/out})$. How the boundary locations, $r_\text{in/out}$, and the truncation values, $k^\pm_\text{max}$, are selected in practice will be discussed in the algorithm section below. The series coefficients $a^i_k, b^i_k$ are found by substituting the above expansions into the field equations \eqref{eq:metric_pert_FD} and solving for the resulting recursion relations. For brevity we do not repeat these relations here; they can be found in Appendix A of Ref.~\cite{Akcay-GSF-circular}. The recursion relations determine the $a^i_{k>0},b^i_{k>0}$ in terms of the first coefficients $a^i_0,b^i_0$, respectively. By selecting appropriate linearly independent vectors of these leading coefficients we construct a basis of linearly independent solutions that span the solution space for the field equations. For example, for the odd radiative modes we have, once the gauge equations are employed, a solution space with two degrees of freedom, i.e., we must solve for $\hb{9}$ and $\hb{10}$. For the outer homogeneous solutions the two basis are formed by setting $\{a^9_0,a^{10}_0\} =\{1,0\}$ and $\{a^9_0,a^{10}_0\} =\{0,1\}$. Similarly we can repeat this with $\{b^9_0,b^{10}_0\}$ for the inner solutions.

In this work, although analytic solutions are now known \cite{Osburn_etal:2014}, we opt to solve for the even static modes numerically as we already have code to do so. For these modes the numerical boundary conditions take the form
\begin{align}
	\hbm{i}(r_\text{in}) 	&= \sum_{k=k^-_\text{min}}^{k^-_\text{max}} b^i_k(r_\text{in}-2M)^k				,				\\
	\hbp{i}(r_\text{out}) 	&= \sum_{k=k^+_\text{min}}^{k^+_\text{max}} \frac{a^i_k + \bar{a}^i_k \log r_\text{out}}{r_\text{out}^k}	.	\label{eq:static_outer_BCs}
\end{align}
How the truncation values $k^\pm_\text{min}$ are selected and the form of the recursion relations for $a^i_k,\bar{a}^i_k,b^i_k$ is again given in Ref.~\cite{Akcay-GSF-circular}. The $\log$ term in Eq.~\eqref{eq:static_outer_BCs} is added to ensure the recursion relations have sufficient degrees of freedom to span the space of solutions to the field equations.

\subsection{Numerical algorithm}

The following steps describe how we calculate the self-force in practice via our frequency-domain effective-source approach.
\begin{enumerate}
	\item{Choose a radial grid to store the values of various fields on. In general we require high resolution near the particle, and lower resolution far from the particle. Though our chosen window function is not formally compact, within our numerical procedure it is effectively compact. It is inside this effectively compact region that we need high resolution. In general we choose our window function to be effectively zero outside the region $(r_0 - 2M, r_0 + 2M)$. Inside this region we find a grid spacing of $M/10$ sufficient (we pick the grid so that it includes $r=r_0$). Outside the (effective) support of the window function we use a grid spacing of $2M$.}
	\item{For radiative modes ($m\neq0$) and even static modes ($\ell=\text{even}\ge2,m=0$) construct numerical boundary conditions at $r=r_\text{out}$ and $r=r_\text{in}$ using the recursion relations in Appendix A of Ref.~\cite{Akcay-GSF-circular}. For each $\ell m$-mode there will be $n_f$ inhomogeneous fields to solve for -- see Table \ref{table:hierachical_structure} -- and correspondingly $n_f$ sets of boundary conditions for the homogeneous fields will be constructed as described in Sec.~\ref{sec:num_BCs}.}
	\item{For a given $\ell m$-mode solve for each basis of homogeneous solutions and store the values of the fields and their derivatives on the preselected radial grid points.}
	\item{For the odd static ($\ell=\text{odd},m=0$) modes and the monopole ($\ell=m=0$) the values of the (in)homogeneous fields and their derivatives can be computed analytically. See Appendix \ref{apdx:monopole} for an explicit overview of the calculation for the monopole mode}.
	\item{For the given $\ell m$-mode compute the effective-source vector and store the results on the radial grid.}
	\item{At each grid point invert the matrix of homogeneous solutions defined in Eq.~\eqref{eq:Phi_matrix} (formed from the previous stored results) and multiply it by the source vector to form the integrand of Eq.~\eqref{eq:Cres}. Store the resulting values of the weighting coefficient integrands at each point on the radial grid.}
	\item{Interpolate the weighting coefficient integrands using standard cubic spline techniques. Numerically integrate the integrand as described by Eq.~\eqref{eq:Cres}. The regular radial metric perturbation fields and their radial derivative are then computed via Eqs.~\eqref{eq:hres} and \eqref{eq:dhres}, respectively.}
	\item{The gauge fields are then constructed as discussed at the end of Sec.~\ref{sec:construct_hres} following the hierarchical structure given in Table \ref{table:hierachical_structure}.}
	\item{The metric and its derivatives are constructed using the formulae given in Appendix \ref{apdx:metric_reconstruction}. The self-force is then constructed via Eq.~\eqref{eq:F_self}.}
\end{enumerate}

For comparison we also compute the retarded field with the method described in Sec.~\ref{sec:ret_solution}. The first four steps of the algorithm in this case are the same as those above. Then for the fifth step we use Eq.~\eqref{eq:ret_weighting_coeffs} to construct the retarded-field weighting coefficients. This step only requires knowledge of the homogeneous fields and the sources given in Appendix \ref{apdx:field_eqs_coupling}. The retarded solutions are then constructed using Eq.~\eqref{eq:hret_construction}. Finally we compute the self-force with both the standard mode-sum prescription described in Sec.~\ref{sec:mode_sum}, which relies on projecting the retarded tensor modes onto a basis of spherical-harmonics before regularization, and the tensor mode-sum prescription we present in Sec.~\ref{sec:mode_sum_tensor}. Note that only the radial component of the self-force requires regularization as it is the only component with non-zero regularization parameters (see Eq.~\ref{eq:mode_sum_reg_params}). Correspondingly, the contributions to both the $t$- and $\varphi$-components of the self-force converge exponentially in both $\ellh$ and $\ell$ (the $\theta$-component is zero by symmetry).

\subsection{Results}

\begin{figure}[htb!]
	\hskip-0.3cm
	\includegraphics[width=8.5cm]{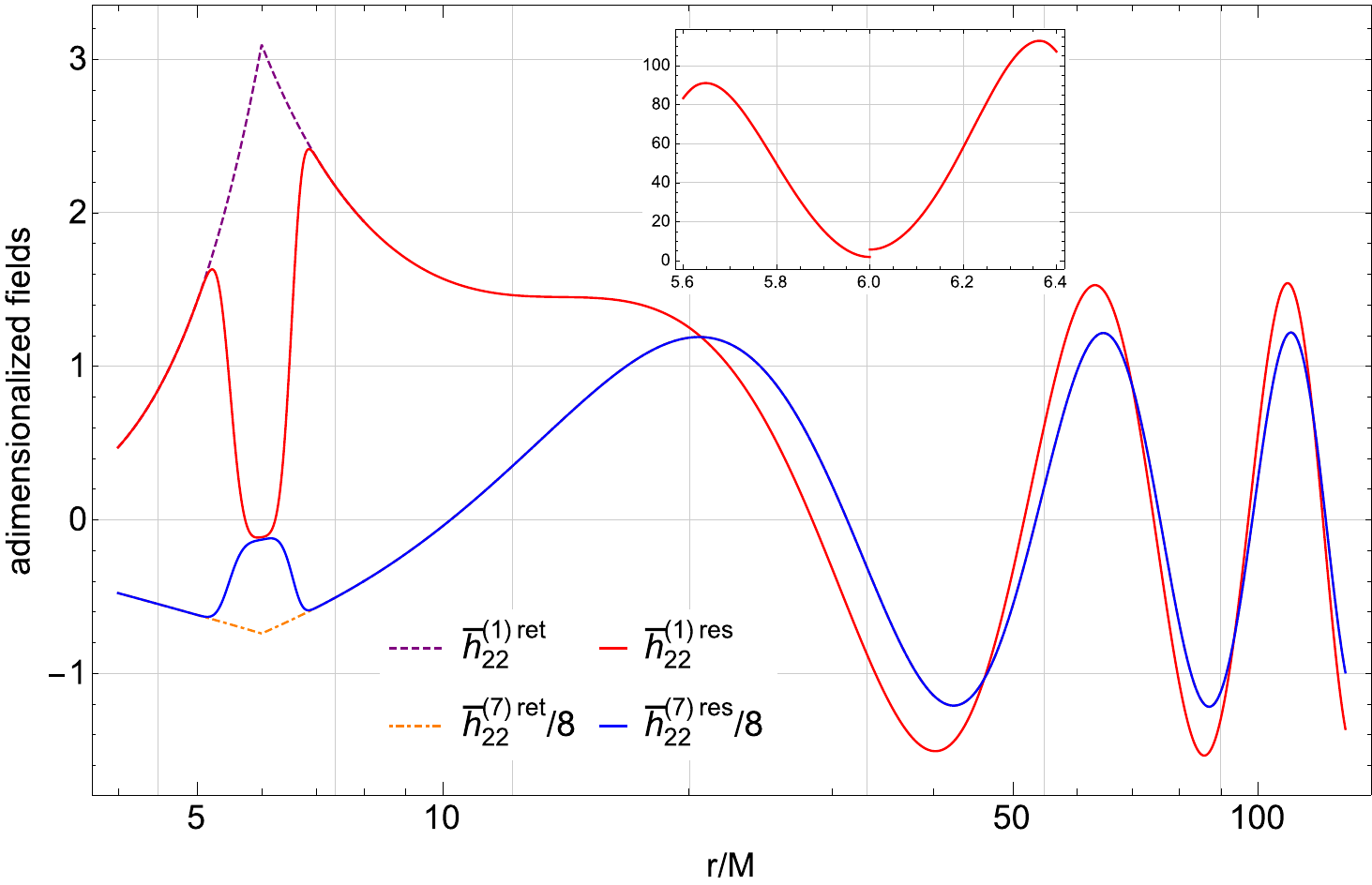}
	\caption{Sample results for the $\ell=2$, $m=2$ mode for a particle orbiting at
$r_0=6M$. Shown are the $\hb{1}$ and (scaled) $\hb{7}$ metric perturbations for
both the residual and retarded fields. At $r=6M$, the upper solid (red) curve shows the
residual field $\bar{h}^{(1)\text{res}}$. The dashed (purple) curve shows the
retarded field $\bar{h}^{(1)\text{ret}}$. Far from the particle the two
coincide. Similarly, the lower solid (blue) curve shows
$\bar{h}^{(7)\text{res}}$ and the dot-dashed (orange) curve shows
$\bar{h}^{(7)\text{ret}}$. The inset shows $\bar{h}^{(1)\text{res}}_{,rr}$ near
the particle. With the punctures we present in the main text the residual fields are $C^1$ at the particle
and correspondingly, as the inset shows, the second radial derivatives of the residual field are discontinuous.
As the residual fields are $C^1$ at the
particle  the self-force can be directly computed from their derivatives.
The punctures we provide online \cite{online} give $C^2$ residual fields at the particle which acts to improve the
convergence rate of the mode-sum, as shown in Fig.~\ref{fig:convergence}.
}\label{fig:residual_field}
\end{figure}

Using the above algorithm we can compute the residual metric perturbation at
$r=r_0$. Using the residual field at the particle we can compute $h^R_{uu}$ and
we find that our results agree with previously published results to relative
accuracy of $10^{-7}$. Taking a radial derivative of the residual metric
perturbation at the particle we can compute the radial self-force without any
further regularization required. For the radial self-force we find agreement
with the previous published results to a relative accuracy of $10^{-6}$ -- see
Table \ref{table:results}. In Fig.~\ref{fig:residual_field} we plot the
residual field for the $(\ell,m,i)=(2,2,1)$ field for a particle orbiting at
$r_0=6M$.

A key feature of our procedure is that we only ever work with tensor-harmonic
modes in constructing the self-force. This is in contrast to the standard
mode-sum scheme whereby the tensor-harmonics of the retarded field are
projected onto a basis of scalar harmonics before regularization. This
projection, though straightforward, is cumbersome to implement (see, e.g.,
Appendix C in Ref.~\cite{Barack-Sago-circular}). Furthermore the coupling
between the tensor and scalar modes means that in practice to calculate
$\ellh_\text{max}$ scalar modes one needs to calculate $\ell_\text{max} =
\ellh_\text{max} + 3$ tensor modes. With our prescription this is not necessary.

\begin{figure}[htb!]
	\includegraphics[width=8.5cm]{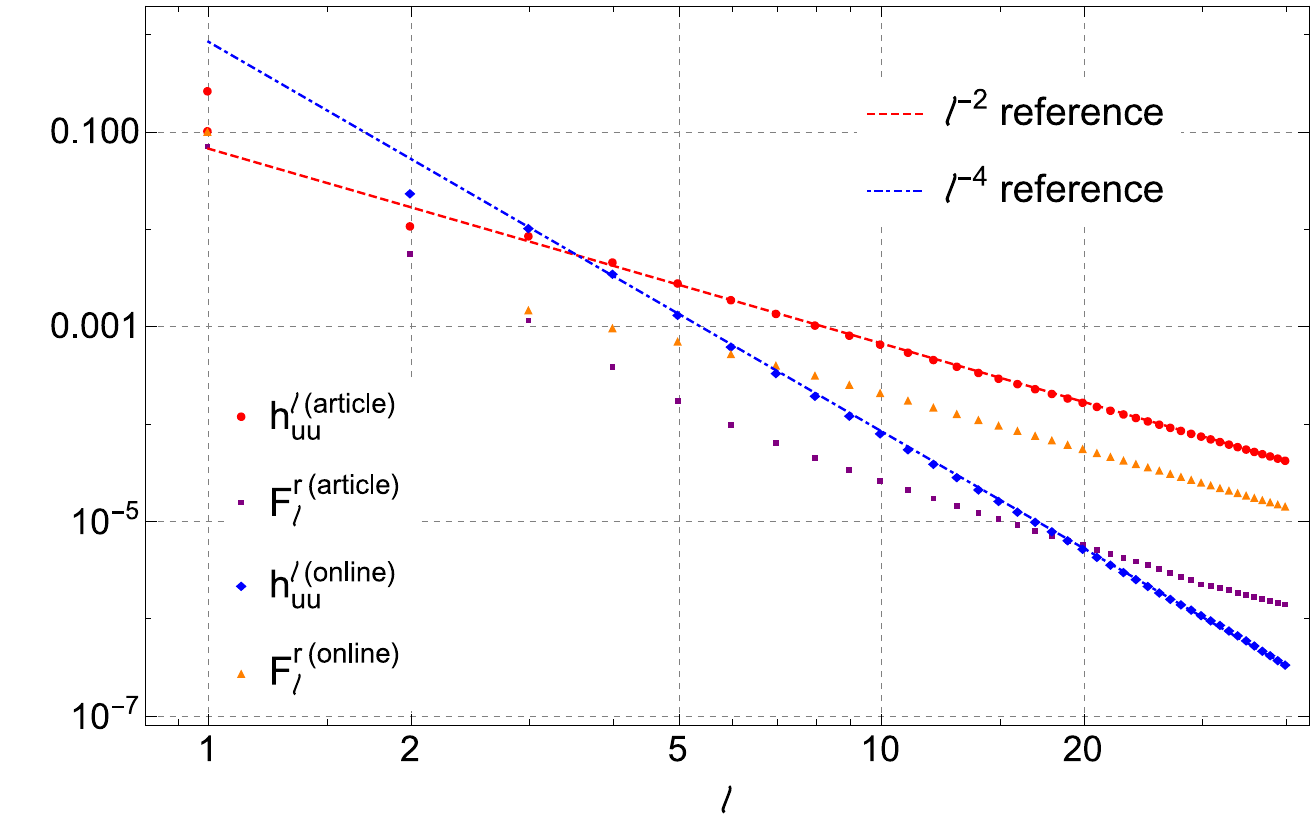}
	\caption{Convergence of the {\it tensor} $\ell$-mode contributions to (the
adimensionalized) $h_{uu}$ and $F^r$ for a particle orbiting at $r_0=6M$. The
punctures we present in this article in Appendix~\ref{apdx:punctures} are
sufficiently regular that the contributions to $h_{uu}$ and $F^r$ drop off as
$\ell^{-2}$. Online \cite{online} we give higher-order punctures that improve the rate of
convergence for $h_{uu}$ to $\ell^{-4}$. This is not the case for $F^r$; the
higher-order punctures are still one order lower than would be required to
improve its rate of convergence. Instead, they merely change the coefficient of
the $\ell^{-2}$ behaviour in such a way that the infinite sum over $\ell$ is
unaffected.}
\label{fig:convergence}
\end{figure}

In Fig.~\ref{fig:convergence} we show the convergence of the tensor mode-sum
for the regular contributions to $h^R_{uu}$ and $F^r$. The punctures we use in
this work are sufficiently regular that, for high-$\ell$ the contributions to
$h^R_{uu}$ and $F^r$ drop off as $\ell^{-4}$ and $\ell^{-2}$, respectively.

In the next section we show how, by taking the limit to the worldline in our
effective-source procedure, we can formulate a tensor-mode mode-sum scheme.

\begin{table}[htb!]
	\footnotesize
\centering
\begin{tabularx}{\columnwidth}{c c c c c}
\toprule
			& $r_0/M$	& this work 					& Akcay \etal~\cite{Akcay-Warburton-Barack,Akcay_etal:2012}	& rel.~diff.		\\
\hline
$h^R_{uu}$	& 6			& $-1.0471852(4)$				& $-1.0471854796(1)$				  &	  	$2\times10^{-7}$						  \\
$F^r$		& 6			& $2.4466487(8)\times10^{-2}$	& $2.4466495(4)\times10^{-2}$	  &		$3\times10^{-7}$						\\
\hline
$h^R_{uu}$	& 10		& $-0.48925802(2)$					& $-0.48925800172(4)$				&	$4\times10^{-8}$						\\
$F^r$		& 10		& $1.3389466(3)\times10^{-2}$		& $1.3389465(7)\times10^{-2}$		&	$3\times10^{-8}$						\\
\toprule
\end{tabularx}
\caption{
Sample results for orbits with $r_0=6M$ and $r_0=10M$ computed with
$\ell_\text{max}=40$. The relative difference between our results and previously
published data is small, being always less than $3\times10^{-7}$. These results
were computed using the higher-order punctures we provide online \cite{online} and by
numerically fitting for the higher-order regularization parameters in order to
speed up the convergence of the sum over tensor $\ell$-modes. Numbers in brackets
denote the estimated error in the final digit of the corresponding result. Note
that the results in this table have been adimensionalized, i.e., $h^R_{uu}$ here
$\equiv (M/\mu)h^R_{uu}$ and $F^r$ here $\equiv (M/\mu)^2
F^r$}\label{table:results}
\end{table}

\section{Mode-sum regularization with tensor-harmonic modes}
\label{sec:mode_sum_tensor}

In addition to their use in the effective source approach described here, the
puncture fields may also be used to improve the efficiency of the traditional
mode-sum scheme. In the standard mode-sum prescription, regularization is achieved
through mode-sum formulae such as Eqs.~\eqref{eq:huu_mode_sum} and \eqref{eq:Fr_mode_sum}, which take the form
\begin{equation}
  \label{eq:msr-scalar}
  F^{\rm R}_{\mu} = \sum_{\ellh=0}^{\infty} \Big[ F^{\ellh\, {\rm ret}}_{\mu} - (2\ellh+1) F^{[-1]}_{\mu} - F^{[0]}_{\mu} \Big] + D_{\mu},
\end{equation}
in the case of the self-force where, recall, an $\ellh$ sub/superscript denotes 
the scalar-harmonic multipole contribution (summed over m). The regularization parameters $F^{[-1]}_{\mu}$
and $F^{[0]}_{\mu}$ are analytically-derived functions of the instantaneous
worldline. Provided the Detweiler-Whiting singular field is used in their
derivation (and the retarded-field modes $F^{\ellh\, {\rm ret}}_{\mu}$ are
in Lorenz gauge), the $D_\mu$ term vanishes. Each term in the sum goes like
$\ellh^{-2}$ and so the partial sums converge as $\ell_{\rm max}^{-1}$.
One can derive additional higher-order regularization parameters to accelerate
this rate of convergence. For example, by subtracting an appropriate term of the
form $\tfrac{F^{[2]}_{\mu}}{(2\ellh-1)(2\ellh+3)}$ one finds that the
terms in the sum now fall off as $\ellh^{-4}$. One can continue in this way
to higher orders, where the order $\ellh^{-n}$ term has the form
\begin{equation}
\label{eq:scalar-rp-ldep}
  \sum_{\ellh=0}^\infty \tfrac{2\ellh+1}{(2\ellh-n+1)(2\ellh-n+3)\ldots(2\ellh+n-1)(2\ellh+n+1)} = 0,
\end{equation}
for $n$ even. The fact that the infinite sum over $\ellh$ vanishes is important as
it guarantees that the subtraction of higher-order regularization parameters
does not affect the numerical result (other than accelerating the rate of
convergence); equivalently, these higher-order terms can be seen to come from
pieces of the Detweiler-Whiting singular field which vanish when evaluated on
the worldline.

In the scalar-field case, this mode-sum formula is a natural choice
as one can choose to work with \emph{scalar} spherical harmonics labeled by
$\ellh$ when solving the field equations. In the gravitational case, the \emph{tensor} spherical harmonics are a
more natural choice and a numerical calculation typically produces tensor
harmonic modes (labeled by $\ell$) for the retarded field. Despite this fact,
existing calculations have relied on a scalar-harmonic mode-sum formula of the
form given in Eq.~\eqref{eq:msr-scalar}. As a result, a necessary step in the
regularization procedure is the projection of the tensor-harmonic modes,
$F^{\ell\, {\rm ret}}_{\mu}$ onto scalar harmonic modes $F^{\ellh\, {\rm
ret}}_{\mu}$. This is undesirable for at least two reasons: (i) the projection
involves cumbersome mode coupling formulas which have to be derived on a
case-by-case basis, and (ii) a given scalar-harmonic mode $\ellh$ couples to
several tensor-harmonic modes (up to $\ellh \pm 2$ for the metric and higher for
some of its derivatives). This second point means that in order to obtain a
given number of scalar $\ellh$ modes, one actually has to compute several higher
tensor-harmonic $\ell$ modes of the retarded field, and these are then lost during the
projection. Given that the cost of computing a given retarded-field mode grows
quadratically with $\ell$, this turns out to be quite a significant increase in
computational cost.

Fortunately, it turns out that the projection onto scalar harmonics is
unnecessary; in this section we will derive tensor-harmonic regularization
parameters which completely eliminate the need for scalar harmonics. This
addresses both issues mentioned above and produces a much simpler, more accurate
and computationally efficient result.

First, we consider what form a tensor-harmonic mode-sum scheme should take. The
$\ell$-dependence of the term of order $\ell^{-n}$ will be given by
\begin{equation}
\label{eq:tensor-rp-ldep}
  \Lambda_{m,n} \equiv \frac{2^{n-2m} (2 \ell+1)(\ell-m+1)_{2 m}}{(2 \ell-2 m+n+1) \left(\ell-m+\frac{n}{2}+\frac{3}{2}\right)_{2 m-n}}
\end{equation}
for $m\ge0$ and $n$ integers. Here, we use the standard notation $(a)_n = a(a+1)\ldots(a+n-1)$ for the
Pochhammer symbol. When $m=0$ we can see that this reduces to the
scalar-harmonic case, Eq.~\eqref{eq:scalar-rp-ldep}, as expected. For $m>0$ and
$n\ge2$ the infinite sum over $\ell$ of any of these terms is zero, meaning we
are free to add them without modifying the final result (other than
accelerating convergence). In Eq.~\eqref{eq:scalar-rp-ldep} this was only true
when the sum starts at $\ell=0$, which would not be appropriate for tensor
harmonics. For our generalized expression, \eqref{eq:tensor-rp-ldep}, this
holds for the sum starting at any value in the range $0 \le \ell \le m$. In
practice we will have $m$ equal to the value of $m'$ used in the punctures and
$n$ will be determined by the power of $\rho$ appearing in the singular field.

Examining the puncture fields in Appendix \ref{apdx:punctures} we can see they are already written in a form where
this $\ell$-dependence is manifestly apparent. The task of producing
tensor-harmonic regularization parameters is therefore merely a matter of
reconstructing the $\ell$ modes of the singular metric perturbation using the
expressions given in Appendix~\ref{apdx:metric_reconstruction}, summing over
$m$ and then evaluating on the worldline. It is most convenient to do so in the
$(\alpha,\beta)$ coordinate system, i.e., using the punctures without the
Wigner-D rotation matrices and evaluating at $\alpha=0$, as then only a small
number of $m'$ modes must be summed over. The only caveat is that this yields
the components of the metric in the $(\alpha,\beta)$ coordinates. We must 
therefore also include a factor of the Jacobian from $(\alpha,\beta)$ to
$(\theta,\varphi)$ coordinates. This Jacobian is given by
\begin{subequations}
\begin{eqnarray}
  \frac{\partial \alpha}{\partial \theta} &=& \frac{- \cos \alpha \sin \beta}{\sqrt{1-\sin^2\alpha \sin^2 \beta}} \approx 0,
\\
  \frac{\partial \alpha}{\partial \varphi} &=& \cos \beta \approx 1,
\\
  \frac{\partial \beta}{\partial \theta} &=& \frac{- \cos \beta}{\sin\alpha\sqrt{1-\sin^2\alpha \sin^2 \beta}} \approx -\frac{1}{\sin \alpha},
\\
  \frac{\partial \beta}{\partial \phi} &=& \frac{- \cos \alpha \sin \beta}{\sin\alpha} \approx 0.
\end{eqnarray}
\end{subequations}
Note that because of the factor of $\tfrac{1}{\sin\alpha}$ appearing here, it is
important to multiply by the Jacobian before taking the limit $\alpha \to 0$.

Since we are interested in computing the metric perturbation and its
derivative (for the self-force), we require mode-sum formulae for all
components of the metric perturbation and its derivative, i.e.,
\begin{equation}
  h^{\rm R}_{\mu \nu} = \sum_{\ell=0}^{\infty} \Big[ h^{\ell\, {\rm ret}}_{\mu \nu} - h^{[0]}_{\mu \nu} \Big]
\end{equation}
and
\begin{equation}
  h^{\rm R}_{\mu \nu, \gamma} = \sum_{\ell=0}^{\infty} \Big[ h^{\ell\, {\rm ret}}_{\mu \nu, \gamma} - (2\ell+1) h^{[-1]}_{\mu \nu, \gamma} - h^{[0]}_{\mu \nu, \gamma} \Big].
\end{equation}
Here, the retarded-field $\ell$ modes are computed in the usual way from
numerical data in the $(\theta,\varphi)$ coordinates,
\begin{equation}
  h^{\ell\, {\rm ret}}_{\mu \nu} \equiv \sum_{m=-\ell}^{\ell} h^{\ell m}_{\mu\nu}\big(r_0, \tfrac{\pi}{2}, 0\big),
\end{equation}
where the $h^{\ell m}_{\mu\nu}$ are constructed by combining the
$h^{(i)}_{\ell m}$ with the spherical harmonics; explicit expressions are given
in Appendix~\ref{apdx:metric_reconstruction}.
The regularization parameters for the metric perturbation are computed using
\begin{equation}
  h^{[0]}_{\mu \nu} = \sum_{m'=-\ell}^{\ell} \Big[ \tfrac{\partial x^{\mu'}}{\partial x^{\mu}}\tfrac{\partial x^{\nu'}}{\partial x^{\nu}} h^{{\rm P}\, \ell m'}_{\mu'\nu'} \Big]_{x = x_0},
\end{equation}
where $x_0$ denotes the point on the worldline, i.e. $r = r_0$ and $\alpha = 0
= \beta$. The only non-zero contributions in our case come from $m'=0$ in the
scalar sector ($i=1,3,6$), $m'=\pm1$ in the vector sector ($i=4,8$), and
$m'=\pm2$ in the tensor sector ($i=7,10$). The regularization parameters for
the radial derivative of the metric perturbation are computed using
\begin{equation}
  \label{eq:modesum-tensor-rderiv}
  h^{[0]}_{\mu \nu,r} =  \sum_{m'=-\ell}^{\ell} \Big[ \tfrac{\partial x^{\mu'}}{\partial x^{\mu}}\tfrac{\partial x^{\nu'}}{\partial x^{\nu}} \partial_{r}h^{{\rm P}\,\ell m'}_{\mu'\nu'} \Big]_{x = x_0},
\end{equation}
where, again, the only non-zero contributions in this case come from $m'=0$ in the
scalar sector ($i=1,3,6$), $m'=\pm1$ in the vector sector ($i=4,8$), and
$m'=\pm2$ in the tensor sector ($i=7,10$). Finally, the regularization
parameters for the $\varphi$ derivative of the metric perturbation are computed
using
\begin{equation}
  h^{[0]}_{\mu \nu,\varphi} = \sum_{m'=-\ell}^{\ell} \bigg[ \tfrac{\partial \alpha}{\partial \varphi} \partial_{\alpha} \Big(\tfrac{\partial x^{\mu'}}{\partial x^{\mu}}\tfrac{\partial x^{\nu'}}{\partial x^{\nu}} h^{{\rm P}\,\ell m'}_{\mu'\nu'} \Big)\bigg]_{x = x_0},
\end{equation}
where the only non-zero contributions in this case come from $m'=\pm1$
for $h_{tr}^{P\, \ell m'}$ (i.e., $i=2$) and $m'=(0,\pm2)$ for
$h_{rA}^{{\rm P}\, \ell m'}$ (i.e., $i=5,9$). Evaluating these with the punctures
given in Appendix~\ref{apdx:punctures} yields the following tensor-harmonic
regularization parameters:
\begin{subequations}
\begin{eqnarray}
  h^{[0]}_{tt} &=& \frac{4 (r_0 - M) \mathcal{K}}{\pi\, r_0^2}\sqrt{\frac{r_0 - 2M}{r_0 - 3M}},
\\
  h^{[0]}_{t\varphi} &=& - \frac{32 M^{1/2} \mathcal{K}}{\pi \, r_0^{1/2}} \sqrt{\frac{r_0 - 2M}{r_0 - 3M}} \Lambda_1,\quad
\\
  h^{[0]}_{rr} &=& \frac{4 \mathcal{K}}{\pi}\frac{(r_0 - 3M)^{1/2}}{(r_0 - 2M)^{3/2}},
\\
  h^{[0]}_{\theta\theta} &=& \frac{4 r_0 \mathcal{K}}{\pi} \sqrt{\frac{r_0 - 2M}{r_0 - 3M}}
  \nonumber \\
  && \quad- \frac{64 M r_0 \mathcal{K}}{\pi (r_0 - 2M)^{1/2}(r_0 - 3M)^{1/2}} \Lambda_2,
\\
  h^{[0]}_{\varphi\varphi} &=& \frac{4 r_0 \mathcal{K}}{\pi} \sqrt{\frac{r_0 - 2M}{r_0 - 3M}}
  \nonumber \\
  && \quad+ \frac{64 M r_0 \mathcal{K}}{\pi (r_0 - 2M)^{1/2}(r_0 - 3M)^{1/2}} \Lambda_2,
\\
  h^{[-1]}_{tt,r} &=& \mp\frac{(r_0 - M)}{r_0^{5/2} (r_0 - 3 M)^{1/2}},
\\
  h^{[0]}_{tt,r} &=& \frac{2 (r_0 - M) [(r_0 - 2 M) \mathcal{E} - 2 (r_0 - 4 M) \mathcal{K}]}{\pi r_0^3 (r_0 - 3 M)^{1/2} (r_0 - 2 M)^{1/2}},
\\
  h^{[-1]}_{rr,r} &=& \mp\frac{(r_0 - 3 M)^{1/2}}{r_0^{1/2} (r_0 - 2 M)^2},
\\
  h^{[0]}_{rr,r} &=& \frac{2 (r_0 - 3 M)^{1/2} [(r_0 - 2 M) \mathcal{E} - 2 r_0 \mathcal{K}]}{\pi r_0 (r_0 - 2 M)^{5/2}},
\\
  h^{[-1]}_{t\varphi,r} &=& \pm \bigg[\frac{2 M^{1/2}}{r_0 (r_0 - 3 M)^{1/2}}\bigg]_{\ell\ge1},
\\
  h^{[0]}_{t\varphi,r} &=& -\frac{16 M ^{1/2} [(r_0 - 2 M) \mathcal{E} + 2 M \mathcal{K}]}{\pi r_0^{3/2} (r_0 - 3 M)^{1/2} (r_0 - 2 M)^{1/2}} \Lambda_1,
\\
  h^{[-1]}_{\varphi\varphi,r} &=& \mp \sqrt{\frac{r_0}{r_0-3M}} \mp \bigg[\frac{M r_0^{1/2}}{(r_0-2M)(r_0-3M)^{1/2}}\bigg]_{\ell\ge2}, \nonumber \\
\\
  h^{[0]}_{\varphi\varphi,r} &=& \frac{2 (\mathcal{E} + 2 \mathcal{K})}{\pi} \sqrt{\frac{r_0 - 2 M}{r_0 - 3 M}}
  \nonumber \\
  && \quad + \frac{32 M (\mathcal{E} + 2 \mathcal{K})}{\pi (r_0 - 3 M)^{1/2} (r_0 - 2 M)^{1/2}} \Lambda_2,
\\
  h^{[-1]}_{\theta\theta,r} &=& \mp \sqrt{\frac{r_0}{r_0-3M}} \pm \bigg[ \frac{M r_0^{1/2}}{(r_0-2M)(r_0-3M)^{1/2}} \bigg]_{\ell\ge2}, \nonumber \\
\\
  h^{[0]}_{\theta\theta,r} &=& \frac{2 (\mathcal{E} + 2 \mathcal{K})}{\pi} \sqrt{\frac{r_0 - 2 M}{r_0 - 3 M}}
  \nonumber \\
  && \quad- \frac{32 M (\mathcal{E} + 2 \mathcal{K})}{\pi (r_0 - 3 M)^{1/2} (r_0 - 2 M)^{1/2}} \Lambda_2,
\\
  h^{[0]}_{tr,\varphi} &=& - \frac{32((r_0 - 2 M) \mathcal{E} - (r_0 - 3 M) \mathcal{K})}{\pi M^{1/2} r_0^{3/2}} \sqrt{\frac{r_0 - 2 M}{r_0 - 3 M}} \Lambda_1,\nonumber \\
\\
  h^{[0]}_{r\varphi,\varphi} &=& \frac{16[(r_0 - 2 M) \mathcal{E} - (r_0 - 3 M) \mathcal{K}]}{\pi (r_0 - 2 M)^{1/2} (r_0 - 3 M)^{1/2}} (\Lambda_1 + 4 \Lambda_2),
  \nonumber \\
\end{eqnarray}
\end{subequations}
where, recall, $\Lambda_1$ and $\Lambda_2$ are given by
Eqs.~\eqref{eq:Lambda_1} and \eqref{eq:Lambda_2}, respectively, and where we
have indicated with a subscript the cases ($h^{[-1]}_{t\varphi,r}$,
$h^{[-1]}_{\theta\theta,r}$ and $h^{[-1]}_{\varphi\varphi,r}$) where a term is
only non-zero above some minimum value of $\ell$. In all of the above
equations, to simplify the presentation we have omitted an overall factor of
the small mass $\mu$.

Finally, we note that these expressions can be combined to produce
tensor-harmonic regularization parameters for the redshift invariant
$h_{\mu\nu} u^\mu u^\nu$ and the radial component of the self-force. Doing so,
we find
\begin{eqnarray}
  H^{[0]} &=& \frac{4\mu}{\pi\sqrt{r_0^2 +\ang_0^2}}\mathcal{K} \nonumber \\
   && - \frac{1}{(2\ell-1)(2\ell+3)}\frac{8 \mu M (6r_0-17M) \mathcal{K}}{\pi r_0 (r_0-3M)^{3/2} (r_0-2M)^{1/2}} \nonumber \\
   && + \frac{1}{(2\ell-3)(2\ell-1)(2\ell+3)(2\ell+5)} \times \nonumber \\
   && \qquad \frac{420 \mu M^2\mathcal{K}}{\pi r_0 (r_0-3M)^{3/2} (r_0-2M)^{1/2}},
\end{eqnarray}
\begin{eqnarray}
  F^{r\pm}_{[-1]} &=& \mp \frac{\mu^2}{2 r_0^2}\left(1-\frac{3M}{r_0}\right)^{1/2}
    \pm \bigg[\frac{2 \mu^2 M (2 M-r_0)}{r_0^{5/2} (r_0-3 M)^{3/2}}\bigg]_{\ell<1} \nonumber \\
    && \qquad \pm \bigg[\frac{\mu^2 M^2}{2 r_0^{5/2} (r_0-3 M)^{3/2}}\bigg]_{\ell<2},
\end{eqnarray}
\begin{eqnarray}
  F^r_{[0]} &=& \frac{\mu^2 r_0 \en_0^2}{\pi(\ang_0^2 + r_0^2)^{3/2}}\left[\mathcal{E} - 2\mathcal{K}\right] \nonumber \\
   && - \frac{1}{(2\ell-1)(2\ell+3)} \times \nonumber \\
   && \qquad \frac{2 \mu^2 M (r_0-2M)^{1/2} [(6r_0-17M) \mathcal{E} + 2 M \mathcal{K}]}{\pi r_0^3 (r_0-3M)^{3/2}} \nonumber \\
   && + \frac{1}{(2\ell-3)(2\ell-1)(2\ell+3)(2\ell+5)} \times \nonumber \\
   && \qquad \frac{105 \mu^2 M^2 (r_0-2M)^{1/2} (\mathcal{E}+2\mathcal{K})}{\pi r_0^3 (r_0-3M)^{3/2}}.
\end{eqnarray}
Note that in giving these parameters we have rewritten $\Lambda_1$ and
$\Lambda_2$ in a form which highlights the fact that $H^{[0]}$ and $F^r_{[0]}$ both match their
scalar-harmonic counterparts, Eq.~\eqref{eq:mode_sum_reg_params}, with the
exception of higher-order terms in $1/\ell$. Since these terms vanish when
summed from $\ell=0$ to infinity they have no impact on the final result and
can be ignored in practice. Importantly, this is not the case for
$F^{r\pm}_{[-1]}$ which differs from its scalar-harmonic version. The
difference is in the presence of the second and third terms, and arises from
the fact our mode sum expression, Eq.~\eqref{eq:modesum-tensor-rderiv}, starts
at $\ell=0$ while it should start at $\ell=1$ for the vector sector and at
$\ell=2$ for the tensor sector. However,
since this term has different limits on either side of the worldline, it
vanishes upon averaging the left and right radial limits. As such, we see that in this
case regularization can be achieved without projection onto scalar harmonics,
by using the scalar-harmonic regularization parameters combined with
an averaging procedure.

\section{Concluding remarks}

In this work we have developed a frequency-domain application of the effective
source approach to computing the self-force on a point mass in a curved
background spacetime. This new method builds on previous work which studied the
case of a scalar-field toy model \cite{Warburton-Wardell}, extending it to the
more physically relevant gravitational case.

With a numerical implementation for the case of a circular orbit in Schwarzschild
spacetime, our results demonstrate that the method can reliably produce accurate
numerical results for the regularized metric perturbation with modest effort and
computational cost. While this is not particularly important in a first order
calculation --- the traditional mode-sum method, for example, can already
produce comparable results with similar, or better, computational efficiency ---
the primary goal of our approach is to develop a set of methods which will be
applicable to a second-order self-force calculation. Our results provide two
key components in that regard:
\begin{enumerate}
  \item Our numerical scheme for solving the sourced field equations in the
        frequency domain carries over immediately to second order. The only
        change will be that the source will be a more complicated function
        involving the first-order metric perturbation.
  \item The source for the second-order field equations is most efficiently
        written in terms of the first order Detweiler-Whiting regular field in
        an extended region near the worldline. Such an approximation is exactly
        the output from our first-order calculation.
\end{enumerate}

In addition to addressing several important aspects of a second-order
self-force calculation, the tensor-harmonic regularization parameters we derived
in Sec.~\ref{sec:mode_sum_tensor} can also be used to improve the computational
efficiency of a first-order mode-sum self-force calculation by avoiding the
need for a cumbersome and wasteful projection onto scalar harmonics. It is
interesting to note the close relation between the tensor harmonic
regularization parameters and those one would obtain using a scalar-harmonic
decomposition. In particular, provided one computes an average of either side
of the (radial) limit to the worldline, we have found that scalar-harmonic
regularization parameters may be used in place of their tensor-harmonic
counterparts. We anticipate that this is more than merely a coincidence - in a
future work we will investigate whether a similar result holds in more general
cases.

There are several future directions in which our results may be extended. Most
important is the application of our approach to the calculation of conservative
effects from the second order gravitational self-force
\cite{Pound:2nd_order,Pound:2014,Pound-Miller:2014}. In addition to this,
it may be interesting to study extensions of the approach beyond circular
orbits, to the Kerr spacetime and to radiation and Regge-Wheeler gauges. With
a view to identifying other important second-order effects, it may also be
interesting to incorporate our method into an orbital evolution scheme which
makes use of a two-timescale expansion of the equations of motion \cite{Hinderer-Flanagan}. Such a
scheme would likely provide a compelling balance of computational efficiency
and faithfulness to the underlying physics of the EMRI problem.

\begin{acknowledgments}

We thank Adam Pound, Leor Barack, Jeremy Miller, Adrian Ottewill and Michael
Boyle for helpful conversations and suggestions.
This material is based upon work supported by the National Science Foundation
under Grant Number 1417132. B.W. was supported by Science Foundation Ireland
under Grant No.~10/RFP/PHY2847, by the John Templeton Foundation New Frontiers
Program under Grant No.~37426 (University of Chicago) - FP050136-B (Cornell
University), and by the Irish Research Council, which is funded under the
National Development Plan for Ireland.
N.W. gratefully acknowledges support from a Marie Curie International Outgoing
Fellowship (PIOF-GA-2012-627781) and the Irish Research Council, which is
funded under the National Development Plan for Ireland.

\end{acknowledgments}

\appendix

\section{Functions on the two-sphere}
\label{sec:harmonics}

When dealing with functions on the two-sphere, there are a wide number of possible conventions. Our
conventions, which are consistent with those of Mathematica \cite{Mathematica} are summarized in
this Appendix.

\subsection{Scalar spherical harmonics}
The associated Legendre polynomials may be defined in terms of derivatives of the standard Legendre
polynomials,
\begin{subequations}
\begin{eqnarray}
  P_\ell^{m}(x) &=& (-1)^m\ (1-x^2)^{m/2}\ \frac{d^m}{dx^m} P_\ell(x) \quad [m \ge 0], \nonumber \\ \\
  P_\ell^{-m}(x) &=& (-1)^m \frac{(\ell-m)!}{(\ell+m)!} P_\ell^{m}(x),
\end{eqnarray}
\end{subequations}
where we have included the Condon-Shortley phase factor $(-1)^m$ and where the Legendre polynomials
satisfy the Legendre equation
\begin{equation}
  (1-x^2) \frac{d^2 P_\ell(x)}{dx^2} - 2 x \frac{dP_\ell(x)}{dx} + \ell(\ell+1)P_\ell(x) = 0.
\end{equation}
We now define the scalar spherical harmonics as
\begin{equation}
   Y_{\ell m}( \theta , \varphi ) = \sqrt{\frac{(2\ell+1)}{4\pi}\frac{(\ell-m)!}{(\ell+m)!}}  \, P_\ell^m ( \cos{\theta} ) \, e^{i m \varphi },
\end{equation}
where $-\ell \le m \le \ell$. Note that $\ell$ and $m$ are merely labels and we will raise and
lower their position freely to wherever they get in the way the least. The scalar spherical
harmonics are orthonormal,
\begin{equation}
 \int_{0}^{2\pi} \int_{0}^\pi Y_{\ell m}( \theta , \varphi ) \, Y_{\ell' m'}^{*}( \theta , \varphi ) \, d\Omega=\delta_{\ell\ell'}\, \delta_{mm'},
\end{equation}
where $d\Omega \equiv \sin\theta d\theta d\varphi$. They also satisfy
\begin{equation}
  Y_{\ell m}^{*}( \theta , \varphi ) = (-1)^m Y_{\ell, -m}( \theta , \varphi ),
\end{equation}
and the completeness relation
\begin{equation}
  \sum_{\ell=0}^{\infty} \sum_{m=-\ell}^{\ell} Y_{\ell m}( \theta , \varphi ) \, Y_{\ell m}^{*}( \theta' , \varphi' )
    = \delta(\cos\theta - \cos\theta') \delta(\varphi - \varphi').
\end{equation}
Since the scalar harmonics form an orthonormal basis, an arbitrary scalar function can be expanded
in spherical harmonics,
\begin{equation} \label{eq:ScalarExpansion}
  f(\theta, \varphi) = \sum_{\ell=0}^{\infty} \sum_{m=-\ell}^{\ell} f_{\ell m} \, Y_{\ell m}( \theta , \varphi ),
\end{equation}
where the coefficients are given by
\begin{equation}
  f_{\ell m} = \int_{0}^{2\pi} \int_{0}^\pi f( \theta , \varphi ) \, Y_{\ell m}^{*}( \theta , \varphi ) \, d\Omega.
\end{equation}
At the pole, $\theta = 0$, only the $m=0$ spherical harmonics are non-zero and
(\ref{eq:ScalarExpansion}) becomes
\begin{equation}
  f(0, \varphi) = \sum_{\ell=0}^{\infty} \sqrt{\frac{2\ell+1}{4\pi}} f_{\ell 0},
\end{equation}
where
\begin{equation}
  f_{\ell 0} = \sqrt{\frac{2\ell +1}{4\pi}} \int_{0}^{2\pi} \int_{0}^\pi f( \alpha , \beta ) \, P_{\ell}(\cos \alpha) \, d\Omega.
\end{equation}
Similarly, the $\theta$ derivative only requires modes $m=-1,1$:
\begin{equation}
  (\partial_\theta f)(0, \varphi) =
    \sum_{\ell=0}^{\infty} \sqrt{\frac{\ell(\ell+1)(2\ell+1)}{16\pi}} (e^{-i \varphi} f_{\ell,-1} - e^{i \varphi} f_{\ell 1}),
\end{equation}
the second $\theta$ derivative requires $m=0,\pm2$:
\begin{eqnarray}
  &&(\partial_{\theta\theta} f)(0, \varphi) = \sum_{\ell=0}^{\infty} \sqrt{\frac{(2\ell+1)}{64\pi}} \bigg[- 2 \ell(\ell+1) f_{\ell 0}
  \nonumber \\
  && \quad + \sqrt{(\ell-1)\ell(\ell+1)(\ell+1)}(e^{-2 i \beta} f_{\ell,-2} + e^{2 i \beta} f_{\ell 2})\bigg], \nonumber \\
\end{eqnarray}
and so on; for $n$ derivatives with respect to $\theta$  we need modes
$m=-n, -n+2, \cdots, n-2, n$.

\subsection{Vector spherical harmonics}
The vector spherical harmonics fall into two categories, those of even parity
($\ell+m$ even) and those of odd parity ($\ell+m$ odd). These categories
reflect a difference in behaviour under the parity operation
$(\theta,\varphi)\rightarrow(\pi-\theta,\varphi+\pi)$; the even-parity
harmonics are invariant under this transformation while the odd-parity
harmonics change sign.

The even-parity vector harmonics are defined by
\begin{equation}
  Z_A^{\ell m} = [\ell(\ell+1)]^{-1/2} D_A Y^{\ell m},
\end{equation}
and the odd-parity harmonics are defined by
\begin{equation}
  X_A^{\ell m} = - [\ell(\ell+1)]^{-1/2} \epsilon_A{}^B D_B Y^{\ell m},
\end{equation}
where $D_A$ is the covariant derivative and $\epsilon_{AB}$ is the Levi-Civita
tensor associated with the metric $\Omega_{AB} = \text{diag}(1,\sin^2\theta)$
on the two-sphere (i.e. $\epsilon_{\theta\varphi} = \sin\theta$,
$\epsilon_{\varphi\theta} = -\sin\theta$,
$\epsilon_{\theta\theta}=0=\epsilon_{\varphi\varphi}$). Explicitly, the
components of the vector harmonics are
\begin{eqnarray}
  Z_\theta^{\ell m} &=& [\ell(\ell+1)]^{-1/2} \partial_\theta Y^{\ell m}, \nonumber \\
    X_\theta^{\ell m} &=& -[\ell(\ell+1)]^{-1/2} \frac{1}{\sin\theta} \partial_\varphi Y^{\ell m}, \nonumber \\
  Z_\varphi^{\ell m} &=& [\ell(\ell+1)]^{-1/2} \partial_\varphi Y^{\ell m}, \nonumber \\
   X_\varphi^{\ell m} &=& [\ell(\ell+1)]^{-1/2} \sin\theta\, \partial_\theta Y^{\ell m}.
\end{eqnarray}
The vector harmonics satisfy the orthonormality relations
\begin{subequations}
\begin{eqnarray}
  \int_{0}^{2\pi} \int_{0}^\pi X_A^{\ell m}( \theta , \varphi ) \, X_{\ell' m'}^{A*}( \theta , \varphi ) \, d\Omega&=&\delta_{\ell\ell'}\, \delta_{mm'}, \qquad \\
  \int_{0}^{2\pi} \int_{0}^\pi Z_A^{\ell m}( \theta , \varphi ) \, Z_{\ell' m'}^{A*}( \theta , \varphi ) \, d\Omega&=&\delta_{\ell\ell'}\, \delta_{mm'}, \qquad \\
  \int_{0}^{2\pi} \int_{0}^\pi X_A^{\ell m}( \theta , \varphi ) \, Z_{\ell' m'}^{A*}( \theta , \varphi ) \, d\Omega&=&0.
\end{eqnarray}
\end{subequations}
They also satisfy
\begin{subequations}
\begin{eqnarray}
  X_A^{\ell m*}( \theta , \varphi ) = (-1)^m X_A^{\ell, -m}( \theta , \varphi ), \\
  Z_A^{\ell m*}( \theta , \varphi ) = (-1)^m Z_A^{\ell, -m}( \theta , \varphi ).
\end{eqnarray}
\end{subequations}
These definitions are consistent with Ref.~\cite{Thorne:1980} and with Ref.~\cite{Martel:2005ir} (apart from
the inclusion of the prefactor $[\ell(\ell+1)]^{-1/2}$ which ensures orthonormality) and relate
to those of Ref.~\cite{Haas:2011np} through the conversion
$Z_A^{\ell m} \to [\ell(\ell+1)]^{1/2} Y_A^{\ell m}$,
$X_A^{\ell m} \to [\ell(\ell+1)]^{1/2} X_A^{\ell m}$.

\subsection{Tensor spherical harmonics}
The tensor spherical harmonics again fall into two categories, those of even parity ($\ell+m$ even) and
those of odd parity ($\ell+m$ odd). The even-parity tensor harmonics are defined by
\begin{equation}
  Z_{AB}^{\ell m} = \left[2\frac{(\ell-2)!}{(\ell+2)!}\right]^{\frac{1}{2}} \big[D_A D_B +\frac{1}{2} \ell(\ell+1) \Omega_{AB} \big] Y^{\ell m},
\end{equation}
and the odd-parity harmonics are defined by
\begin{equation}
  X_{AB}^{\ell m} = - \left[2\frac{(\ell-2)!}{(\ell+2)!}\right]^{\frac{1}{2}} \epsilon_{(A}{}^C D_{B)} D_C Y^{\ell m}.
\end{equation}
Explicitly, the components of the tensor harmonics are
\begin{subequations}
\begin{eqnarray}
  Z_{\theta\theta}^{\ell m} &=& \Bigg[2\frac{(\ell-2)!}{(\ell+2)!}\Bigg]^{\frac{1}{2}}\Big[\partial_{\theta\theta} + \tfrac{1}{2}\ell(\ell+1)\Big] Y^{\ell m},
\\
  Z_{\theta\varphi}^{\ell m} &=& \Bigg[2\frac{(\ell-2)!}{(\ell+2)!}\Bigg]^{\frac{1}{2}}\Big[\partial_{\theta\varphi} - \cot\theta \, \partial_\varphi\Big] Y^{\ell m},
\\
  Z_{\varphi\varphi}^{\ell m} &=& \Bigg[2\frac{(\ell-2)!}{(\ell+2)!}\Bigg]^{\frac{1}{2}}\Big[\partial_{\varphi\varphi} + \sin\theta \cos\theta\, \partial_\theta
  \nonumber \\
  && \qquad \qquad \qquad \qquad + \tfrac{1}{2}\ell(\ell+1)\sin^2\theta\Big]  Y^{\ell m}, \qquad\quad
\\
  X_{\theta\theta}^{\ell m} &=& - \Bigg[2\frac{(\ell-2)!}{(\ell+2)!}\Bigg]^{\frac{1}{2}}\frac{1}{\sin\theta}\Big[\partial_{\theta\varphi} - \cot\theta \, \partial_\varphi\Big] Y^{\ell m},
\\
  X_{\theta\varphi}^{\ell m} &=& - \Bigg[2\frac{(\ell-2)!}{(\ell+2)!}\Bigg]^{\frac{1}{2}}\frac{1}{2\sin\theta} \Big[ \partial_{\varphi\varphi} - \sin^2\theta\, \partial_{\theta\theta},
  \nonumber \\
  && \qquad \qquad \qquad \qquad  + \sin\theta \cos\theta\, \partial_\theta \Big]Y^{\ell m}
\\
  X_{\varphi\varphi}^{\ell m} &=& \Bigg[2\frac{(\ell-2)!}{(\ell+2)!}\Bigg]^{\frac{1}{2}}\sin\theta\Big[\partial_{\theta\varphi} - \cot\theta\, \partial_\varphi\Big] Y^{\ell m}. \qquad\quad
\end{eqnarray}
\end{subequations}
The tensor harmonics satisfy the orthonormality relations
\begin{subequations}
\begin{eqnarray}
  \int_{0}^{2\pi} \int_{0}^\pi X_{AB}^{\ell m}( \theta , \varphi ) \, X_{\ell' m'}^{AB*}( \theta , \varphi ) \, d\Omega&=& \delta_{\ell\ell'}\, \delta_{mm'}, \qquad
\\
  \int_{0}^{2\pi} \int_{0}^\pi Z_{AB}^{\ell m}( \theta , \varphi ) \, Z_{\ell' m'}^{AB*}( \theta , \varphi ) \, d\Omega&=& \delta_{\ell\ell'}\, \delta_{mm'}, \qquad
\\
  \int_{0}^{2\pi} \int_{0}^\pi X_{AB}^{\ell m}( \theta , \varphi ) \, Z_{\ell' m'}^{AB*}( \theta , \varphi ) \, d\Omega&=&0,
\end{eqnarray}
\end{subequations}
and the identity
\begin{equation}
  \Omega^{AB} Z_{AB}^{\ell m} = 0 = \Omega^{AB} X_{AB}^{\ell m}.
\end{equation}
They also satisfy
\begin{subequations}
\begin{eqnarray}
  X_{AB}^{\ell m*}( \theta , \varphi ) = (-1)^m X_{AB}^{\ell, -m}( \theta , \varphi ), \\
  Z_{AB}^{\ell m*}( \theta , \varphi ) = (-1)^m Z_{AB}^{\ell, -m}( \theta , \varphi ).
\end{eqnarray}
\end{subequations}
These definitions are consistent with Thorne \cite{Thorne:1980} and relate to those of
Ref.~\cite{Martel:2005ir} through an orthonormality factor,
$Z_{AB}^{\ell m} \to \left[2\frac{(\ell-2)!}{(\ell+2)!}\right]^{1/2} Y_{AB}^{\ell m}$,
$X_{AB}^{\ell m} \to \left[2\frac{(\ell-2)!}{(\ell+2)!}\right]^{1/2} X_{AB}^{\ell m}$.

\subsection{Rotations}
Under a rotation of the coordinate system which is represented by the Euler angles $\alpha, \beta,
\gamma$, the spherical harmonic components transform according to
\begin{equation}
  \label{eq:rot-harmonic}
  f_{\ell m} (\theta, \varphi) = \sum_{m'=-\ell}^{\ell} D_{mm'}^{\ell} (\alpha, \beta, \gamma) f_{\ell m'}  (\theta', \varphi'),
\end{equation}
where $D_{mm'}^{\ell} (\alpha, \beta, \gamma)$ is the Wigner-D matrix
\cite{Wigner}. Here, we use the convention that the Euler angles correspond to
a $z-y-z$ counterclockwise rotation and our convention\footnote{This convention
is different from that of \emph{Mathematica} \cite{Mathematica} and Wigner
\cite{Wigner}. Our $D_{mm'}^{\ell} (\alpha, \beta, \gamma)$ is related to
theirs by a change in the signs of $m$ and $m'$ \cite{Rose}.} for
$D_{mm'}^{\ell} (\alpha, \beta, \gamma)$ is consistent with Rose \cite{Rose}.
Using these conventions, the Wigner-D matrix satisfies
\begin{equation}
  D_{m_1 m_2}^{\ell}(\alpha, \beta, \gamma) = e^{-i m_1 \alpha - i m_2 \gamma} D_{m_1 m_2}^{\ell} (0, \beta, 0).
\end{equation}
The vector and tensor harmonics also transform in a similar way \cite{Challinor:2000df}, i.e.,
\begin{subequations}
\begin{eqnarray}
  &&X^{\ell m}_{A} (\theta, \varphi) = \nonumber \\
  && \quad \frac{\partial x^{A'}}{\partial x^{A}}\sum_{m'=-\ell}^{\ell}  D_{mm'}^{\ell} (\alpha, \beta, \gamma) X^{\ell m'}_{A'} (\theta', \varphi'), \qquad \quad\\
  &&Z^{\ell m}_{A} (\theta, \varphi) = \nonumber \\
  && \quad \frac{\partial x^{A'}}{\partial x^{A}}\sum_{m'=-\ell}^{\ell} D_{mm'}^{\ell} (\alpha, \beta, \gamma) Z^{\ell m'}_{A'} (\theta', \varphi'), \qquad \quad\\
  &&X^{\ell m}_{AB} (\theta, \varphi) = \nonumber \\
  && \quad \frac{\partial x^{A'}}{\partial x^{A}} \frac{\partial x^{B'}}{\partial x^{B}} \sum_{m'=-\ell}^{\ell} D_{mm'}^{\ell} (\alpha, \beta, \gamma) X^{\ell m'}_{A'B'}  (\theta', \varphi'), \qquad \quad\\
  &&Z^{\ell m}_{AB} (\theta, \varphi) = \nonumber \\
  && \quad \frac{\partial x^{A'}}{\partial x^{A}} \frac{\partial x^{B'}}{\partial x^{B}} \sum_{m'=-\ell}^{\ell} D_{mm'}^{\ell} (\alpha, \beta, \gamma) Z^{\ell m'}_{A'B'} (\theta', \varphi'), \qquad \quad
\end{eqnarray}
\end{subequations}
which is equivalent to stating that the vector- and tensor-harmonic components
of a tensor transform according to Eq.~\eqref{eq:rot-harmonic}.
Finally, we note that the Wigner-D matrix relates to the spin-weighted spherical harmonics:
\begin{equation}
  D_{ms}^{\ell}(\alpha, \beta, \gamma) = (-1)^{s} \sqrt{\frac{4\pi}{2\ell+1}} {}_{-s} Y_{\ell m}^* (\beta, \alpha) e^{-i s \gamma},
\end{equation}
which for the spin-0 case gives a relation to the scalar harmonics,
\begin{equation}
  D_{m 0}^{\ell}(\alpha, \beta, 0) = \sqrt{\frac{4\pi}{2\ell+1}} Y_{\ell m}^* (\beta, \alpha).
\end{equation}

\subsection{Tensor-harmonic basis in Schwarzschild spacetime}
Barack and Lousto \cite{Barack-Lousto} used the above bases of scalar, vector and tensor harmonics
to construct a basis of harmonics for the components of a symmetric rank-$2$ tensor $t_{\mu\nu}$
defined on a Schwarzschild background spacetime. This basis was later modified slightly
by Barack and Sago \cite{Barack-Sago-circular} to improve the behaviour of some components near the horizon.
In particular, they chose a basis of ten fields in $t-r$ space defined by
\begin{subequations}
  \label{eq:hbar_decomp}
\begin{eqnarray}
  t^{(1)}_{\ell m} &=& \int_{0}^{2\pi} \int_{0}^\pi r\, (t_{tt}+ f^2 t_{rr}) Y^{\ast}_{\ell m} \, d\Omega, \\
  t^{(2)}_{\ell m} &=& \int_{0}^{2\pi} \int_{0}^\pi 2\, r f\, t_{tr} Y^{\ast}_{\ell m} \, d\Omega, \\
  t^{(3)}_{\ell m} &=& \int_{0}^{2\pi} \int_{0}^\pi rf^{-1}\, (t_{tt}- f^2 t_{rr}) Y^{\ast}_{\ell m} \, d\Omega, \\
  t^{(4)}_{\ell m} &=& \int_{0}^{2\pi} \int_{0}^\pi 2 \big[\ell(\ell+1)\big]^{1/2} t_{tA} Z^{A\ast}_{\ell m} \, d\Omega, \\
  t^{(5)}_{\ell m} &=& \int_{0}^{2\pi} \int_{0}^\pi 2 \big[\ell(\ell+1)\big]^{1/2} f \, t_{rA} Z^{A\ast}_{\ell m} \, d\Omega, \\
  t^{(6)}_{\ell m} &=& \int_{0}^{2\pi} \int_{0}^\pi \frac{1}{r} t_{AB} \Omega^{AB} Y^{\ast}_{\ell m} \, d\Omega, \\
  t^{(7)}_{\ell m} &=& \int_{0}^{2\pi} \int_{0}^\pi \frac{2}{r} \left[2\frac{(\ell-2)!}{(\ell+2)!}\right]^{-1/2} \times \nonumber \\
      && \qquad t_{AB} \Big(Z^{AB\ast}_{\ell m} - \frac12 \Omega^{AB} \Omega_{CD} Z^{CD\ast}_{\ell m} \Big) \, d\Omega, \nonumber \\ \\
  t^{(8)}_{\ell m} &=& -\int_{0}^{2\pi} \int_{0}^\pi 2 \big[\ell(\ell+1)\big]^{1/2} t_{tA} X^{A\ast}_{\ell m} \, d\Omega, \\
  t^{(9)}_{\ell m} &=& -\int_{0}^{2\pi} \int_{0}^\pi 2 \big[\ell(\ell+1)\big]^{1/2} f \, t_{rA} X^{A\ast}_{\ell m} \, d\Omega, \\
  t^{(10)}_{\ell m} &=& - \int_{0}^{2\pi} \int_{0}^\pi \frac{2}{r} \left[2\frac{(\ell-2)!}{(\ell+2)!}\right]^{-1/2} \times \nonumber \\
      && \qquad t_{AB} \Big(X^{AB\ast}_{\ell m} - \frac12 \Omega^{AB} \Omega_{CD} X^{CD\ast}_{\ell m} \Big)\, d\Omega, \nonumber \\
\end{eqnarray}
\end{subequations}
where $f\equiv (1-2M/r)$.
The harmonics $i=1,\ldots,7$ are of even parity, while the harmonics $i=8,9,10$
are of odd parity.

Barack and Sago represented this basis in terms of a set of ten tensors defined by
\begin{subequations}
\begin{eqnarray}
  Y^{(1)}_{\mu\nu} &=& \tfrac{1}{\sqrt{2}} (\delta_\mu^t \delta_\nu^t + f^{-2} \delta_\mu^r \delta_\nu^r )Y^{\ell m}, \\
  Y^{(2)}_{\mu\nu} &=& \tfrac{1}{f\sqrt{2}} (\delta_\mu^t \delta_\nu^r + \delta_\mu^r \delta_\nu^t )Y^{\ell m}, \\
  Y^{(3)}_{\mu\nu} &=& \tfrac{f}{\sqrt{2}} (\delta_\mu^t \delta_\nu^t - f^{-2} \delta_\mu^r \delta_\nu^r )Y^{\ell m}, \\
  Y^{(4)}_{\mu\nu} &=& \tfrac{r}{\sqrt{2}} (\delta_\mu^t Z^{\ell m}_\nu + Z^{\ell m}_\mu \delta_\nu^t ), \\
  Y^{(5)}_{\mu\nu} &=& \tfrac{r}{f\sqrt{2}} (\delta_\mu^r Z^{\ell m}_\nu + Z^{\ell m}_\mu \delta_\nu^r ), \\
  Y^{(6)}_{\mu\nu} &=& \tfrac{r^2}{\sqrt{2}} \Omega_{AB} \delta_\mu^A \delta_\nu^B Y^{\ell m}, \\
  Y^{(7)}_{\mu\nu} &=& r^2 (Z^{\ell m}_{\mu\nu} - \tfrac12 Z^A{}_A \Omega_{\mu\nu}), \\
  Y^{(8)}_{\mu\nu} &=& -\tfrac{r}{\sqrt{2}} (\delta_\mu^t X^{\ell m}_\nu + X^{\ell m}_\mu \delta_\nu^t ), \\
  Y^{(9)}_{\mu\nu} &=& -\tfrac{r}{f\sqrt{2}} (\delta_\mu^r X^{\ell m}_\nu + X^{\ell m}_\mu \delta_\nu^r ), \\
  Y^{(10)}_{\mu\nu} &=& r^2 (X^{\ell m}_{\mu\nu} - \tfrac12 X^A{}_A \Omega_{\mu\nu}).
\end{eqnarray}
\end{subequations}
With the exception of $i=3$, this basis is an orthonormal set in the sense that
\begin{equation}
  \int_{0}^{2\pi} \int_{0}^\pi \eta^{\tau\mu}\eta^{\kappa\nu}Y^{(i) \ell m}_{\mu\nu} \, Y^{(j)\ell' m'*}_{\tau\kappa} \, d\Omega = \delta_{ij} \delta_{\ell\ell'}\, \delta_{mm'},
\end{equation}
where $\eta^{\tau\kappa} \equiv \text{diag}(1,f^2,r^{-2}, r^{-2}\sin^{-2}\theta)$. For
$i=3$, the set is also orthogonal, but $Y^{(3)}_{\mu\nu}$ has a norm of $f^2$.

Finally, we note that Barack and Sago factored out the coefficients
\begin{align}
a^{(i)}_\ell = \frac{1}{\sqrt{2}}
  \left\{
    \begin{array}{ll}
      1, & i = 1,2,3,6, \\
      (\ell(\ell+1))^{-1/2}, & i = 4,5,8,9,\\
      ((\ell-1)\ell(\ell+1)(\ell+2))^{-1/2}, & i = 7,10,
    \end{array}
  \right.
\end{align}
out from the tensor-harmonic fields, $\bar{h}^{(i)}$, in order to make some of
their expressions for, e.g., the field equations more compact. We likewise
use these coefficients in Eqs.~\eqref{eq:hilm-tr} and \eqref{eq:hPilm-tr}.

\subsection{Metric reconstruction}\label{apdx:metric_reconstruction}

Rebuilding the original metric perturbation, $h_{\mu\nu}$, from the $\hb{i}_{\ell m}$
fields is straightforward. The necessary equation can be derived using
Eq.~\eqref{eq:hbar_decomp} along with the fact that a trace reversal,
$h_{\mu\nu} = \bar{h}_{\mu\nu}-\frac{1}{2}g_{\mu\nu}\bar{h}$, is
equivalent to the interchange $h^{(3)} \Leftrightarrow h^{(6)}$. This gives
\begin{equation}
	h_{\mu\nu} = \frac{\mu}{2r}\sum_{\ell}^\infty\sum_{m=-\ell}^\ell h_{\mu\nu}^{\ell m}e^{-i\omega_m t},
\end{equation}
where
\begin{eqnarray}
h^{\ell m}_{tt} &=& \left(\bar{h}_{\ell m}^{(1)} + f(r) \bar{h}_{\ell m}^{(6)}\right)Y^{\ell m}\label{eq:h_tt},\\
h^{\ell m}_{tr} &=& f(r)^{-1}\bar{h}_{\ell m}^{(2)}Y^{\ell m}\label{eq:h_tr},\\
h^{\ell m}_{rr} &=& f(r)^{-2}\left(\bar{h}_{\ell m}^{(1)} - f\bar{h}_{\ell m}^{(6)}\right)Y^{\ell m}\label{eq:h_rr},\\
h^{\ell m}_{tA} &=& \frac{r}{\sqrt{\ell(\ell+1)}} \left(\bar{h}_{\ell m}^{(4)} Z_A^{\ell m} - \bar{h}_{\ell m}^{(8)} X_A^{\ell m}\right) \label{eq:h_tA},\\
h^{\ell m}_{rA} &=& \frac{r}{f(r)\sqrt{\ell(\ell+1)}} \left(\bar{h}_{\ell m}^{(5)} Z_A^{\ell m} - \bar{h}_{\ell m}^{(9)} X_A^{\ell m}\right) \label{eq:h_rA},\\
h^{\ell m}_{AB} &=& r^2 \Omega_{AB} \bar{h}^{(3)}_{\ell m} Y^{\ell m} \nonumber \\
   && \quad + r^2 \sqrt{2\frac{(\ell-2)!}{(\ell+2)!}} \Big(\bar{h}_{\ell m}^{(7)} (Z^{\ell m}_{AB} - \tfrac12 Z^C{}_{C} \Omega_{AB}) \nonumber \\
   && \qquad - \bar{h}_{\ell m}^{(10)} (X^{\ell m}_{AB} - \tfrac12 X^C{}_{C} \Omega_{AB})\Big), \qquad \label{eq:h_AB}
\end{eqnarray}
and where the sum over $\ell$ begins at $\ell=0$ for the scalar sector
(i.e., $\bar{h}^{1}_{\ell m}$, $\bar{h}^{2}_{\ell m}$, $\bar{h}^{3}_{\ell m}$ and
$\bar{h}^{6}_{\ell m}$), at $\ell=1$ for the vector sector
(i.e., $\bar{h}^{4}_{\ell m}$, $\bar{h}^{5}_{\ell m}$, $\bar{h}^{8}_{\ell m}$ and
$\bar{h}^{9}_{\ell m}$), and at $\ell=2$ for the tensor sector
(i.e., $\bar{h}^{7}_{\ell m}$ and $\bar{h}^{10}_{\ell m}$).

\begin{widetext}
\section{Field equations and retarded field sources}\label{apdx:field_eqs_coupling}
The coupling terms in the frequency-domain field equation
\eqref{eq:metric_pert_FD} are given by
\begin{align}
	\mathcal{M}^{(1)}{}_{(j)} \hb{j} & =  \f{M}{r^2} f \hb{3}_{,r_\ast} + \f{f}{2r^2}\left(1-\f{4M}{r}\right) \left(\hb{1}-\hb{5} - f \hb{3} \right) - \f{f^2}{2r^2} 	\left(1-\f{6M}{r}\right) \hb{6}, \label{eq:eq_R1} \\
	\mathcal{M}^{(2)}{}_{(j)} \hb{j} & =  \f{1}{2}ff' \hb{3}_{,r_\ast} + \f{1}{2} f'  \left[ i \omega \left(\hb{1}-\hb{2}\right) + \hb{2}_{,r_\ast} - \hb{1}_{,r_\ast} \right] + 	\f{f^2}{2 r^2} \left(\hb{2}-\hb{4}\right) \nn \\ & - \f{f f'}{2r}\left( \hb{1} - \hb{5} - f \hb{3} - 2f \hb{6} \right), \label{eq:eq_R2} \\
	\mathcal{M}^{(3)}{}_{(j)} \hb{j} & =  - \f{f}{2r^2} \left[\hb{1} - \hb{5} - \left(1-\f{4M}{r}\right) \left(\hb{3} + \hb{6}\right) \right], \label{eq:eq_R3} \\
	\mathcal{M}^{(4)}{}_{(j)} \hb{j} & =  \f{1}{4} f' \left[ i \omega \left(\hb{5}-\hb{4}\right) + \hb{4}_{,r_\ast} - \hb{5}_{,r_\ast} \right] - \f{1}{2} \ell (\ell+1) \f{f}{ r^2} 	\hb{2} \nn \\ & - \f{f f'}{4r}\left( 3\hb{4} + 2\hb{5} - \hb{7} + \ell (\ell+1) \hb{6} \right)  , \label{eq:eq_R4} \\
	\mathcal{M}^{(5)}{}_{(j)} \hb{j} & =  \f{f}{r^2} \left[ \left(1-\f{9M}{2r}\right) \hb{5} - \f{\ell(\ell+1)}{2}\left(\hb{1} - f \hb{3} \right) + \f{1}{2} \left(1-\f{3M}{r}\right) 	\left( \ell(\ell+1) \hb{6} - \hb{7} \right) \right], \label{eq:eq_R5} \\
	\mathcal{M}^{(6)}{}_{(j)} \hb{j} & =  - \f{f}{2r^2} \left[\hb{1} - \hb{5} - \left(1-\f{4M}{r}\right) \left(\hb{3} + \hb{6}\right) \right], \label{eq:eq_R6} \\
	\mathcal{M}^{(7)}{}_{(j)} \hb{j} & =  - \f{f}{2r^2} \left(\hb{7} + \lambda \hb{5} \right), \label{eq:eq_R7} \\
	\mathcal{M}^{(8)}{}_{(j)} \hb{j} & = \f{1}{4} f' \left[ i \omega \left(\hb{9}-\hb{8}\right) + \hb{8}_{,r_\ast} - \hb{9}_{,r_\ast} \right] -  \f{f f'}{4r}\left( 3\hb{8} + 	2\hb{9} - \hb{10}  \right) , \label{eq:eq_R8} \\
	\mathcal{M}^{(9)}{}_{(j)} \hb{j} & =  \f{f}{r^2}\left(1-\f{9M}{2r}\right) \hb{9} - \f{f}{2r^2}\left(1-\f{3M}{r}\right) \hb{10}, \label{eq:eq_R9} \\  
	\mathcal{M}^{(10)}{}_{(j)}\hb{j} &=  -\frac{f}{2r^2}\left(\bh{10} + \lambda\bh{9}\right).	  \label{eq:eq_R10}
\end{align}
where $\lambda = (\ell-1)(\ell+2)$.

The sources to the field equation \eqref{eq:metric_pert_FD} take the form
\begin{equation}
	\mathcal{J}(r) = -\frac{16\pi\en}{f_0^2} \alpha^{(i)}\delta(r-r_0) \left\{\begin{array}{c} Y^{\ell m*}(\pi/2,\Omega_\varphi t),\;\;i=1,\dots,7\\ Y^{\ell m*}_{,\theta}(\pi/2,\Omega_\varphi t),\;\;i=8,9,10 \end{array}\right.
\end{equation}
where
\begin{align}
  \alpha^{(1)} = f_0^2/r_0, \quad
  \alpha^{(2)} = 0, \quad
  \alpha^{(3)} = f_0/r_0, \quad
  \alpha^{(4)} = 2if_0 m \Omega_\varphi, \quad
  \alpha^{(5)} = 0, \quad
  \nonumber \\
  \alpha^{(6)} = r_0\Omega_\varphi^2, \quad
  \alpha^{(7)} = r_0\Omega_\varphi^2\left[\ell(\ell+1) - 2 m^2\right], \quad
  \alpha^{(8)} = 2f_0\Omega_\varphi, \quad
  \alpha^{(9)} = 0, \quad
  \alpha^{(10)} = 2imr_0\Omega_\varphi^2.
\end{align}

\section{Puncture functions for circular orbits in Lorenz gauge}
\label{apdx:punctures}
In this Appendix, we give our explicit expressions for the Lorenz-gauge puncture
fields, $\bar{h}^{(i)P}_{\ell m}$, for the case of a circular geodesic orbit in
Schwarzschild spacetime. These punctures contain all pieces of the
Detweiler-Whiting singular field necessary to compute the regularized components
of the metric and its first derivatives. Written as tensor-harmonic modes
in the $(\theta,\varphi)$ coordinate system, the punctures are given by
\begin{eqnarray}
\bar{h}_{\ell m}^{(1)\rm{P}} &=&
   r D^\ell_{m,0} \sqrt{\frac{4\pi}{2\ell+1}} \Bigg[
      \frac{8(r_0-2M)^{3/2} \mathcal{K}}{\pi r_0^2 (r_0-3M)^{1/2}}
      - (2 \ell+1) |\Delta r| \frac{2 (r_0-2M)}{r_0^{5/2} (r_0-3 M)^{1/2}}
   \nonumber \\ && \qquad
   + \Delta r \frac{4(r_0-2M)^{1/2}[(r_0-2M) \mathcal{E} -2(r_0-4M) \mathcal{K}]}{\pi r_0^3 (r_0-3M)^{1/2}} 
   \Bigg], \label{eq:hP1}
\\
\bar{h}_{\ell m}^{(2)\rm{P}} &=&
   r f(r) [D^\ell_{m,1}-D^\ell_{m,-1}] \sqrt{\frac{4\pi}{2\ell+1}}\sqrt{\frac{1}{\ell(\ell+1)}} \Bigg[
      \frac{64(r_0-2M)^{1/2}[(r_0-2M)\mathcal{E} - (r_0-3M) \mathcal{K}]}{\pi r_0^{3/2} M^{1/2} (r_0-3M)^{1/2}} \Lambda_1
   \Bigg], \label{eq:hP2}
\\
\bar{h}_{\ell m}^{(3)\rm{P}} &=&
   \frac{r}{f(r)} D^\ell_{m,0} \sqrt{\frac{4\pi}{2\ell+1}} \Bigg[
      \frac{8(r_0-2M)^{3/2} \mathcal{K}}{\pi r_0^2 (r_0-3M)^{1/2}}
      - (2 \ell+1) |\Delta r| \frac{2 (r_0-2M)}{r_0^{5/2} (r_0-3 M)^{1/2}}
   \nonumber \\ && \qquad
      + \Delta r \frac{4(r_0-2M)^{1/2}[(r_0-2M) \mathcal{E} -2(r_0-4M) \mathcal{K}]}{\pi r_0^3 (r_0-3M)^{1/2}}
   \Bigg], \label{eq:hP3}
\\
\bar{h}_{\ell m}^{(4)\rm{P}} &=&
   \ell(\ell+1) [D^\ell_{m,1}-D^\ell_{m,-1}] \sqrt{\frac{4\pi}{2\ell+1}}\sqrt{\frac{1}{\ell(\ell+1)}} \Bigg[
      -\frac{64 (r_0-2 M)^{3/2} (\mathcal{E}-\mathcal{K})}{\pi  M^{1/2} r_0^{1/2} (r_0-3 M)^{1/2}}\Lambda_1
   \nonumber \\ && \qquad
      +\frac{48 (r_0-2 M)^{1/2} [2 (r_0-2 M) \mathcal{E}-(2 r_0-5 M) \mathcal{K}]}{\pi M^{1/2} r_0^{1/2} (r_0-3 M)^{1/2} (2 \ell-1) (2 \ell+3)}
      -(2 \ell+1) |\Delta r| \frac{2 M^{1/2} }{r_0 (r_0-3 M)^{1/2}}
   \nonumber \\ && \qquad
      +\Delta r \bigg(\frac{32 (r_0-2 M)^{1/2} [(r_0-4 M) \mathcal{E}-(r_0-5 M) \mathcal{K}]}{\pi  M^{1/2} r_0^{3/2} (r_0-3 M)^{1/2}} \Lambda_1
   \nonumber \\ && \qquad
        \qquad-\frac{24 [(r_0-2 M) (2 r_0-9 M) \mathcal{E}-2 (11 M^2-7 M r_0+r_0^2) \mathcal{K}]}{\pi M^{1/2} r_0^{3/2} (r_0-3 M)^{1/2} (r_0-2 M)^{1/2} (2 \ell-1) (2 \ell+3)}\bigg)
   \Bigg], \label{eq:hP4}
\\
\bar{h}_{\ell m}^{(5)\rm{P}} &=&
  \ell(\ell+1) f(r) [D^\ell_{m,2}+D^\ell_{m,-2}] \sqrt{\frac{4\pi}{2\ell+1}}\sqrt{\frac{1}{(\ell-1)\ell(\ell+1)(\ell+2)}} \times
   \nonumber \\ && \quad
   \Bigg[
      \frac{256 [(4r_0-11M)(r_0-2M)\mathcal{E}-(4r_0-9M)(r_0-3M)\mathcal{K})}{\pi M (r_0-2 M)^{1/2} (r_0-3 M)^{1/2}} \Lambda_2
   \nonumber \\ && \qquad
      - \frac{(\ell-1)(\ell+2)}{(2\ell-3)(2\ell-1)(2\ell+3)(2\ell+5)} \frac{640 [(8r_0-23M)(r_0-2M)\mathcal{E}-(8r_0-19M)(r_0-3M)\mathcal{K})}{\pi M (r_0-2 M)^{1/2} (r_0-3 M)^{1/2}}
   \Bigg]
   \nonumber \\ &&
   + f(r) D^\ell_{m,0} \sqrt{\frac{4\pi}{2\ell+1}} \Bigg[
      \frac{64 [-(r_0-2M)\mathcal{E}+(r_0-3M)\mathcal{K})}{\pi (r_0-2 M)^{1/2} (r_0-3 M)^{1/2}} \Lambda_1
   \Bigg], \label{eq:hP5}
\\
\bar{h}_{\ell m}^{(6)\rm{P}} &=&
   \frac{1}{r} D^\ell_{m,0} \sqrt{\frac{4\pi}{2\ell+1}} \Bigg[
     \frac{8 M r_0 \mathcal{K}}{\pi (r_0-2M)^{1/2} (r_0-3 M)^{1/2}} 
     - (2 \ell+1) |\Delta r| \frac{2 M r_0^{1/2}}{(r_0-2M) (r_0-3 M)^{1/2}}
   \nonumber \\ &&
     \qquad + \Delta r \frac{4 M (\mathcal{E}+2\mathcal{K})}{\pi (r_0-2M)^{1/2} (r_0-3 M)^{1/2}}
  \Bigg], \label{eq:hP6}
\\
\bar{h}_{\ell m}^{(7)\rm{P}} &=&
   \frac{(\ell-1)\ell(\ell+1)(\ell+2)}{r} \Big[D^\ell_{m,-2}+D^\ell_{m,2}\Big] \sqrt{\frac{4\pi}{2\ell+1}} \sqrt{\frac{1}{(\ell-1)\ell(\ell+1)(\ell+2)}} \times \nonumber \\
   && \quad\Bigg[
     - \frac{128 r_0 [4(r_0-2M)(2r_0-5M)\mathcal{E} - (8 r_0^2- 40 M r_0 + 51 M^2) \mathcal{K}]}{3 \pi M(r_0 - 2 M)^{1/2} (r_0 - 3 M)^{1/2}} \Lambda_2 \nonumber \\
   && \qquad
     + \frac{1}{(2\ell-1)(2\ell+3)} \frac{160 r_0 [8(r_0-2M)(2r_0-5M)\mathcal{E} - (4 r_0 - 9 M)(4 r_0 - 11M) \mathcal{K}]}{3 \pi M(r_0 - 2 M)^{1/2} (r_0 - 3 M)^{1/2}} \nonumber \\
   && \qquad
     -(2 \ell+1) |\Delta r| \frac{M r_0^{1/2}}{(r_0-2M)(r_0 - 3 M)^{1/2}} \nonumber \\
   && \qquad
     + \Delta r \bigg(\frac{64 [(8 r_0^2 - 48 M r_0 + 67 M^2)\mathcal{E} - 2(4 r_0^2- 26 M r_0 + 39 M^2) \mathcal{K}]}{3 \pi M(r_0 - 2 M)^{1/2} (r_0 - 3 M)^{1/2}} \Lambda_2 \nonumber \\
   && \qquad
     \qquad - \frac{1}{(2\ell-1)(2\ell+3)} \frac{80 [(16 r_0^2 - 96 M r_0 + 131 M^2)\mathcal{E} - 2(8 r_0^2 - 52 M r_0 + 81 M^2) \mathcal{K}]}{3 \pi M(r_0 - 2 M)^{1/2} (r_0 - 3 M)^{1/2}} \bigg)
  \Bigg], \label{eq:hP7}
\\
\bar{h}_{\ell m}^{(8)\rm{P}} &=&
   i \ell(\ell+1) [D^\ell_{m,1}+D^\ell_{m,-1}] \sqrt{\frac{4\pi}{2\ell+1}}\sqrt{\frac{1}{\ell(\ell+1)}} \Bigg[
      -\frac{64 (r_0-2 M)^{1/2} ((r_0-2M)\mathcal{E}-(r_0-3M)\mathcal{K})}{\pi  M^{1/2} r_0^{1/2} (r_0-3 M)^{1/2}}\Lambda_1
   \nonumber \\
   && \qquad
      +\frac{1}{(2\ell-1)(2\ell+3)} \frac{48 (r_0-2 M)^{1/2} [2 (r_0-2 M) \mathcal{E}-(2 r_0-5 M) \mathcal{K}]}{\pi M^{1/2} r_0^{1/2} (r_0-3 M)^{1/2}}
      +(2 \ell+1) |\Delta r| \frac{2 M^{1/2} }{r_0 (r_0-3 M)^{1/2}}
   \nonumber \\
   && \qquad
      +\Delta r \bigg(\frac{32 [(r_0-2 M)(r_0-5 M) \mathcal{E}-(r_0-3 M) (r_0-4 M) \mathcal{K}]}{\pi  M^{1/2} r_0^{3/2} (r_0-2 M)^{1/2} (r_0-3 M)^{1/2}} \Lambda_1
   \nonumber \\
   && \qquad
        \qquad -\frac{1}{(2\ell-1)(2\ell+3)} \frac{24 [(r_0-2 M) (2 r_0-9 M) \mathcal{E}-2 (11 M^2-7 M r_0+r_0^2) \mathcal{K}]}{\pi M^{1/2} r_0^{3/2} (r_0-3 M)^{1/2} (r_0-2 M)^{1/2}}\bigg)
   \Bigg], \label{eq:hP8}
\\
\bar{h}_{\ell m}^{(9)\rm{P}} &=&
   i \ell(\ell+1)  f(r) [D^\ell_{m,2}-D^\ell_{m,-2}]\sqrt{\frac{4\pi}{2\ell+1}}\sqrt{\frac{1}{(\ell-1)\ell(\ell+1)(\ell+2)}} \times
   \nonumber \\ && \quad
   \Bigg[
      \frac{512 (r_0-3 M)^{1/2} [2(r_0-2M)\mathcal{E}-(2r_0-5M)\mathcal{K})}{\pi M (r_0-2 M)^{1/2}} \Lambda_2
   \nonumber \\ && \qquad
      - \frac{(\ell-1)(\ell+2)}{(2\ell-3)(2\ell-1)(2\ell+3)(2\ell+5)} \frac{640 [(8r_0-23M)(r_0-2M)\mathcal{E}-(8r_0-19M)(r_0-3M)\mathcal{K})}{\pi M (r_0-2 M)^{1/2} (r_0-3 M)^{1/2}}
   \Bigg], \label{eq:hP9}
\\
\bar{h}_{\ell m}^{(10)\rm{P}} &=&
   \frac{i (\ell-1)\ell(\ell+1)(\ell+2)}{r} \Big[D^\ell_{m,-2}-D^\ell_{m,2}\Big] \sqrt{\frac{4\pi}{2\ell+1}} \sqrt{\frac{1}{(\ell-1)\ell(\ell+1)(\ell+2)}} \times \nonumber \\
   && \quad\Bigg[
     - \frac{512 r_0 (r_0 - 2 M)^{1/2} [(2r_0-5M)\mathcal{E} - 2(r_0-3 M) \mathcal{K}]}{3 \pi M (r_0 - 3 M)^{1/2}} \Lambda_2 \nonumber \\
   && \qquad
     + \frac{1}{(2\ell-1)(2\ell+3)} \frac{160 r_0 [8(r_0-2M)(2r_0-5M)\mathcal{E} - (4 r_0 - 9 M)(4 r_0 - 11M) \mathcal{K}]}{3 \pi M(r_0 - 2 M)^{1/2} (r_0 - 3 M)^{1/2}} \nonumber \\
   && \qquad
     +(2 \ell+1) |\Delta r| \frac{M r_0^{1/2}}{(r_0-2M)(r_0 - 3 M)^{1/2}} \nonumber \\
   && \qquad
     + \Delta r \bigg(\frac{256 [2(r_0-4M)(r_0-2M)\mathcal{E} - (r_0-3M)(2r_0-7M) \mathcal{K}]}{3 \pi M(r_0 - 2 M)^{1/2} (r_0 - 3 M)^{1/2}} \Lambda_2 \nonumber \\
   && \qquad
     \qquad - \frac{1}{(2\ell-1)(2\ell+3)} \frac{80 [(16 r_0^2 - 96 M r_0 + 131 M^2)\mathcal{E} - 2(8 r_0^2 - 52 M r_0 + 81 M^2) \mathcal{K}]}{3 \pi M(r_0 - 2 M)^{1/2} (r_0 - 3 M)^{1/2}} \bigg)
  \Bigg], \label{eq:hP10}
\end{eqnarray}
where $D^\ell_{mm'} \equiv D^\ell_{mm'}(\pi,\pi/2,\pi/2)$ is the Wigner-D matrix
corresponding to the rotation from $(\alpha,\beta)$ coordinates to
$(\theta,\varphi)$ coordinates and, recall, $\Lambda_1$ and $\Lambda_2$ are defined in Eqs.~\eqref{eq:Lambda_1} and \eqref{eq:Lambda_2}, respectively.
\end{widetext}

\section{Monopole contribution to the Lorenz-gauge metric perturbation}\label{apdx:monopole}

In this section we calculate the monopole ($\ell=0$) contribution to the
retarded and residual metric perturbation at the particle. There is some subtlety
to Lorenz-gauge monopole perturbations which we will highlight, but we refer the
reader to the given references for more detailed information. We begin by
providing the field equations and basis of homogeneous solutions for the
monopole perturbation.

\subsection{Field equations and basis of homogeneous solutions}

The generic form of the field equations is given by
Eq.~\eqref{eq:metric_pert_FD}. For the monopole perturbation, which has
$l=0,m=0$ and $\omega_m = 0$, only the $(i)=1,3,6$ modes are excited. The field
equations can be further simplified using the gauge equation
\eqref{eq:gauge_even2} to decouple the $\hb{6}$ field --- see discussion around
Table~\ref{table:hierachical_structure}. The remaining field equations for
$\hb{1},\hb{3}$ are given by
\begin{equation}
	\hb{1}_{,rr} = - \frac{1}{r^2 f}\left[(r-4M)\hb{1}_{,r} - \hb{1} - f^2 (r \hb{3}_{,r} - \hb{3})\right], \label{eq:monopole_h1}\\
\end{equation}
\begin{equation}
	\hb{3}_{,rr} = - \frac{1}{r^2}\left[r\hb{3}_{,r} - \hb{3} + \frac{1}{f^2}\left((4M-r)\hb{1}_{,r} + \hb{1}\right)\right]. \label{eq:monopole_h3}
\end{equation}

In order to display the basis of homogeneous solutions, let us define
\begin{align}
	H &\equiv (M/\mu)\left\{h_{tt}, h_{rr}, r^{-2}h_{\theta\theta} = (r\sin\theta)^{-2}h_{\varphi\varphi} \right\}			\nonumber\\
	  &= \frac{M}{4\sqrt{\pi}r}\left\{\hb{1} + f \hb{6}, f^{-2}(\hb{1} - f\hb{6}),\hb{3}\right\}
\end{align}
where the second line derives from the metric reconstruction formulae
\eqref{eq:h_tt}, \eqref{eq:h_rr} and \eqref{eq:h_AB}, noting that
$Y_{00}=\tfrac{1}{2\sqrt{\pi}}$. The inverse relations are
\begin{align}
	\hb{1} &= 2\sqrt{\pi}\mu^{-1}r(h_{tt} + f^2 h_{rr}),  \label{eq:hb1_monopole}\\
	\hb{3} &= 4\sqrt{\pi}\mu^{-1}r^{-1}h_{\theta\theta},  \label{eq:hb3_monopole}\\
	\hb{6} &= 2\sqrt{\pi}\mu^{-1}\frac{r}{f}(h_{tt} - f^2 h_{rr}).
\end{align}

A complete basis of homogeneous solutions to the two coupled monopole field equations \eqref{eq:monopole_h1} and \eqref{eq:monopole_h3} is given by \cite{Akcay-Warburton-Barack}
\begin{align}
	H_A 	=& \left\{ -f, f^{-1},1\right\},																						\label{eq:monopole_H_A}				\\
	H_B 	=& \left\{-\frac{fM}{r^3}P(r), \frac{f^{-1}}{r^3}Q(r), \frac{f}{r^2}P(r) \right\},										\\
	H_C		=& \left\{-\frac{M^4}{r^4}, \frac{M^3 f^{-2}(3M-2r)}{r^4}, \frac{M^3}{r^3}\right\},											\\
	H_D 	=& \left\{ \frac{M}{r^4}\left[W(r)+rP(r)f\ln f - 8M^3 \ln\tfrac{r}{M}\right]\right.	,								\nonumber\\
						& \frac{f^{-2}}{r^4}\left[K(r)-rQ(r) f\ln f - 8M^3(2r-3M)\ln\tfrac{r}{M}\right]	,							\nonumber\\
						& \left. \frac{1}{r^3}\left[3r^3-W(r)-rP(r)f\ln f+8M^3\ln\tfrac{r}{M}\right]\right\},							\label{eq:monopole_H_D}	
\end{align}
where
\begin{align}
  P(r) &= r^2 + 2rM + 4M^2, \\
  Q(r) &= r^3 - r^2M - 2rM^2 +12M^3, \\
  W(r) &= 3r^3 - r^2M - 4rM^2 -28M^3/3, \\
  K(r) &= r^3M - 5r^2M^2 - 20rM^3/3 +28M^4.
\end{align}
By substitution, it is straightforward to verify that the set $\{H_A, H_B, H_C,
H_D\}$ are solutions to the homogeneous field equations \eqref{eq:monopole_h1}
and \eqref{eq:monopole_h3}.

When constructing the inhomogeneous monopole solution it is important to ensure
that it represents a particle with the correct mass-energy. That is, by
Birkhoff's theorem, for the spherically symmetric monopole perturbation the
solution must have the same geometry as the Schwarzschild solution with mass
$M$ for $r<r_0$. For $r>r_0$ the solution must again be that of Schwarzschild
geometry but with mass $M+\mu\en_0$. As we now briefly discuss, perhaps the most
natural method for constructing the inhomogeneous monopole perturbation does
not satisfy this condition.

First we note that the solutions $H_A$ and $H_B$ are regular at the event
horizon but approach nonzero constants as $r\rightarrow\infty$. Conversely,
$H_C$ and $H_D$ are regular at infinity, but are singular on the horizon. It is
therefore tempting to construct the inhomogeneous `internal' solution for $r\le
r_0$ as a weighted sum of the $\{H_A,H_B\}$ basis functions and an `external'
solution for $r \ge r_0$ as a weighted sum of the $\{H_C, H_D\}$ basis
functions. This turns out to not give a solution that has the correct
mass-energy \cite{Dolan-Barack:GSF_m_mode}. Instead a correction term, $\Delta H_\text{ih}$,
must be added \cite{Akcay-Warburton-Barack} so that for the retarded field we have
\begin{equation}\label{eq:monopole_inhom_sol}
	H^\ret_{\ell=0} = \Delta H_\text{ih} + \left\{
     \begin{array}{lr}
       C_A H_A + C_B H_B , 						& r \le r_0,	\\
       C_C H_C + C_D H_D ,						& r \ge r_0,
     \end{array}
   \right.
\end{equation}
where $C_A, C_B, C_C, C_D$ are constant weighting
coefficients and 
\begin{align}\label{eq:DeltaH_ih}
	\Delta H_\text{ih} = -C_A(H_A - H_B).
\end{align}

A curious, but well understood, feature of the Lorenz-gauge monopole
perturbation is that the $tt$ component approaches a nonzero constant at
infinity \cite{Detweiler-Poisson,Detweiler-circular}. When computing gauge-invariant quantities, 
care must be taken to account for this minor peculiarity
of the Lorenz gauge, as discussed in
Refs.~\cite{Sago-Barack-Detweiler,Damour-EOB-SF,Dolan-Barack:GSF_m_mode} (for
circular orbits) and
Refs.~\cite{Barack-Sago-eccentric,Barack-Sago-precession,Akcay_etal:2015} (for
eccentric orbits).

Before we proceed it will be useful to define a matrix of homogeneous solutions by
\begin{align}
	\Phi = \left(\begin{array}{cccc} -\hb{1}_A & -\hb{1}_B 	& \hb{1}_C	& \hb{1}_D	\\
	-\hb{3}_A 		& -\hb{3}_B 		& \hb{3}_C		& \hb{3}_D		\\
	-\hb{1}_{A ,r}	& -\hb{1}_{B,r} 	& \hb{1}_{C,r}	& \hb{1}_{D,r}	\\
	-\hb{3}_{A ,r}	& -\hb{3}_{B,r} 	& \hb{3}_{C,r}	& \hb{3}_{D,r}	\\
	\end{array}\right),
\end{align}
where the matrix elements are constructed by applying the relations
\eqref{eq:hb1_monopole} and \eqref{eq:hb3_monopole} to the basis of homogeneous
solutions \eqref{eq:monopole_H_A}-\eqref{eq:monopole_H_D}.

\subsection{Retarded solution for the monopole mode}

The retarded field at the particle is constructed via Eq.~\eqref{eq:hret_construction} where the sources are given by
\begin{align}
	\mathcal{J}^{(1)} = -\frac{8 \en_0 \sqrt{\pi}}{r_0}, \qquad \mathcal{J}^{(3)} = \frac{\mathcal{J}^{(1)}}{f_0}.
\end{align}
Following Eq.~\eqref{eq:ret_weighting_coeffs} the ($r$-independent) coefficients are given by
\begin{align}\label{eq:monopole_ret_Cs}
	(C_A\,\,C_B\,\,C_C\,\,C_D)^T = \Phi^{-1}.(0\,\,0\,\,\mathcal{J}^{(1)}\,\,\mathcal{J}^{(3)})^T.
\end{align}
Explicitly the weighting coefficients take the form
\begin{align}
	C_A &= -\frac{2M }{\sqrt{r_0(r_0-3M)}},													\\
	C_B &= \frac{8M + (6M - 2r_0)\ln f}{3\sqrt{r_0(r_0-3M)}},								\\
	C_C &= \frac{2\left[8Mr_0 -3r_0^2 - 12M^2+24M(3M-r_0)\ln\tfrac{r_0}{M}\right]}{9M\sqrt{r_0(r_0-3M)}},		\\
	C_D &= \frac{2}{3}\sqrt{1-\frac{3M}{r_0}}.
\end{align}
The monopole contribution to the retarded metric perturbation everywhere in the spacetime is then given by Eq.~\eqref{eq:monopole_inhom_sol}.

\subsection{Residual solution for the monopole mode}

The residual metric perturbation is constructed using Eq.~\eqref{eq:hres} and Eq.~\eqref{eq:Cres}. The effective sources, $S^{(i)\text{eff}}$, that appear in the latter equation are constructed by making the replacements $\hb{1/3} \rightarrow \mathcal{W}\hbP{1/3}$ in Eqs.~\eqref{eq:hb1_monopole} and \eqref{eq:hb3_monopole} for $S^{(1)\text{eff}}$ and $S^{(3)\text{eff}}$, respectively. The two necessary punctures are given by Eqs.~\eqref{eq:hP1} and \eqref{eq:hP3} and $\mathcal{W}$ is the window function. The resulting effective sources are rather cumbersome so we will not display them but they are easily constructed using computer algebra packages.

Following Eq.~\eqref{eq:hres} the ($r$-dependent) weighting coefficients can be solved for via
\begin{align}
	(C_A^\res\,\,C_B^\res\,\,C_C^\res\,\,C_D^\res)^T = \int_a^b\Phi^{-1}(0\,\,0\,\,S^{(1)\text{eff}}\,\,S^{(3)\text{eff}})^T\,dr,
\end{align}
where for $C_A$ and $C_B$ the limits on the integral are $a=r,b=\infty$ and for $C_C$ and $C_D$ the limits are $a=2M,b=r$. The monopole contribution to the residual metric perturbation everywhere in the spacetime is then given by 
\begin{align}
	H^\res_{\ell=0}(r) =& \Delta H_\text{ih} + C_A^\res(r) H_A(r) + C_B(r)^\res H_B(r) \nonumber \\
					 & + C_C(r)^\res H_C(r) + C_D^\res(r) H_D(r).					
\end{align}
Here, although we are calculating the residual field, $\Delta H_\text{ih}$ is still given by Eq.~\eqref{eq:DeltaH_ih} with constant $C_A$ and $C_B$ calculated via Eq.~\eqref{eq:monopole_ret_Cs}.

\bibliographystyle{apsrev4-1}
\bibliography{references}

\end{document}